# Collaborative Distributed Machine Learning

DAVID JIN, Karlsruhe Institute of Technology, Karlsruhe, Germany and KASTEL Security Research Labs, Karlsruhe, Germany

NICLAS KANNENGIESSER, Karlsruhe Institute of Technology, Karlsruhe, Germany and KASTEL Security Research Labs, Karlsruhe, Germany

SASCHA RANK, Karlsruhe Institute of Technology, Karlsruhe, Germany and KASTEL Security Research Labs, Karlsruhe, Germany

ALI SUNYAEV, Technical University of Munich, Campus Heilbronn, Heilbronn, Germany

Various collaborative distributed machine learning (CDML) systems, including federated learning systems and swarm learning systems, with different key traits were developed to leverage resources for the development and use of machine learning (ML) models in a confidentiality-preserving way. To meet use case requirements, suitable CDML systems need to be selected. However, comparison between CDML systems to assess their suitability for use cases is often difficult. To support comparison of CDML systems and introduce scientific and practical audiences to the principal functioning and key traits of CDML systems, this work presents a CDML system conceptualization and CDML archetypes.

## 1 Introduction

Development of machine learning (ML) models requires sufficient training data in terms of quantity and quality to make meaningful predictions with little generalization error. However, sufficient training data is often unavailable to single parties, which can prevent adequate training of ML models [33] and result in inappropriate generalization errors [3]. To reduce generalization errors of ML models, developers request training data from third parties. Training data retrievals from third parties are often subject to compliance and technical challenges [1, 78, 84] that hinder collection of sufficient training data. For example, strict data protection laws and regulations prohibit the disclosure of specific kinds of data, such as personal data by the General Data Protection Regulation of the European Union [23] and organizational data by the Healthcare Insurance Portability and Accountability Act of the USA [24]. Insufficient computing resources inhibit the transfer of large data sets from data centers to developers in acceptable time [78, 84]. To reduce generalization errors of ML models by using training data from multiple parties, an ML paradigm is required that tackles such compliance and technical challenges.

Collaborative distributed ML (CDML) is an ML paradigm that can be implemented to overcome, in particular, compliance and technical challenges related to using data from third parties [8, 45, 87, 90, 97, 105]. In CDML systems, such as federated learning systems [8], split learning systems [90], and swarm learning systems [97], (quasi-)autonomous agents (referred to as agents in the following) train (parts of) ML models on their local training data and self-controlled compute in a distributed manner. Plain training data remains confidential because agents transmit their locally computed training results (i.e., interim results) to other agents, for example, parameter values [53], gradients [63], and activations [90]. Using interim results from other agents, agents improve their local (parts of) ML models. CDML can help solve technical challenges by saving bandwidth. Instead of large amounts of plain training data, only small interim results are transferred between parties.

CDML bears potential for various use cases with different requirements in addition to keeping training data confidential. For instance, language models for effective next-word prediction in virtual smartphone keyboards must be trained on large quantities of heterogeneous and partially sensitive training data representative of user inputs. To access enough training data, CDML systems must be scalable to involve millions [35] or even billions of agents [8]. In contrast, financial institutions use ML models to gain competitive advantages through reliable predictions of financial risks in portfolio management [16, 69]. To protect such competitive advantages, financial institutions keep their ML models confidential. Thus, CDML systems for financial risk predictions must enable collaborative training of ML models without disclosing ML models and training data to competitors.

Specialized CDML systems were developed to meet different use case requirements. For instance, federated learning systems are scalable to engage billions of agents to train ML models for next-word prediction [104]. Split learning systems are unsuitable for this purpose due to the sequential processing of interim results [87]. Conversely, split learning systems seem to be suitable for training ML models to manage financial portfolios because ML model confidentiality is protected in the learning process. Federated learning systems require agents to disclose ML models and, thus, are unsuitable for use cases requiring ML model confidentiality. To design CDML systems that meet use case requirements (e.g., high scalability, ML model confidentiality, and high robustness of the training process), developers need to carefully and systematically compare CDML systems.

Systematic comparison between CDML systems requires thorough understanding of their designs and key traits. Insufficient understanding can lead developers to design CDML systems that fail their purposes, for example, when ML models for portfolio management are inadvertently leaked in unsuitable training processes. However, it is unclear what CDML systems exist and how they differ in their designs because literature on CDML systems is scattered. Moreover, the key traits of CDML system designs remain largely unknown, which makes systematic comparison and informed selection of CDML systems suitable for use cases difficult. A systematization of literature on CDML systems is needed to support comparison and selection of CDML systems for use cases. We ask the following research questions:

RQ1: *What are commonalities and differences between CDML systems?*

RQ2: *What are the key traits of principal CDML system designs?*

We applied a three-step research approach. First, by reviewing publications on CDML, we developed a CDML system conceptualization that specifies the fundamentals of CDML systems (e.g., agent roles and their interactions) and design options for customizing CDML systems (e.g., combinations of agent roles in single agents and communication paths between agents). Second, we analyzed literature on CDML system designs to extract their key traits. Third, we developed CDML archetypes based on the commonalities and differences between the conceptualized CDML systems and extracted key traits.

This work has the following principal contributions to practice and research. In response to RQ1, we present a CDML system conceptualization. The conceptualization introduces the main design commonalities of CDML systems and explains design options for customizing CDML system designs to meet use case requirements. This provides an overview of the CDML paradigm and helps assess its general suitability for use cases. Moreover, by showcasing four CDML archetypes, we inform of combinations of design options commonly used in practice and research. The CDML archetypes can be used as blueprints to tailor CDML archetypes to use cases. In response to RQ2, we present key traits of CDML archetypes. The key traits help better understand how CDML systems can be designed to offer desired key traits to meet use case requirements. By using the CDML archetypes and their key traits, we support better evaluation of CDML system designs in terms of their suitability for use cases before implementing them.

The remainder of this work is structured into six sections. First, we explain the foundations of CDML, outline research trends related to CDML, and describe related research on comparing CDML systems. Second, we describe how we developed the CDML system conceptualization using the Gaia methodology for agent-oriented analysis and design [102]. Moreover, we describe how we developed CDML archetypes and identified their use cases. Third, we present the CDML system conceptualization, including the principal functioning of CDML systems and design options for CDML systems. Fourth, we describe CDML archetypes and explain how different combinations of design options can lead to key traits of CDML systems. Fifth, we discuss our principal findings and describe the contributions and limitations of this work and outline future research directions. Last, we conclude our work with a summary and our takeaways.

## 2 Background and Related Research

To point out usefulness of CDML, the following presents a basic CDML scenario. We explain why common collaborative machine learning (CML) and distributed machine learning (DML) approaches are less suitable for the CDML scenario than CDML. Next, we describe trends related to CDML system development. We offer an overview of related research on comparisons between CDML systems and strengthen the need for the CDML system conceptualization presented in this work.

### 2.1 Basic CDML Scenario

Parties in the basic CDML scenario can be organizations (e.g., in inter-organizational settings), departments (e.g., in intra-organizational settings), and individuals. Party $A$ (e.g., a pharmaceutical company) aims to develop an ML model (e.g., an ML model that predicts the success of medications in treating diseases). However, $A$ lacks sufficient training data and seeks access to more training data to improve the ML model. Parties $B$ (e.g., a hospital) and $C$ (e.g., a biotech company) possess substantial training data that could help improve the ML model of $A$. $B$ and $C$ agree to support $A$ in developing the ML model if the following requirement is met: the training data of each party must not be transmitted to other parties (e.g., due to data protection laws or the financial costs of transmitting large amounts of data), neither in plain text nor encrypted.

*Optional Security Requirements for CDML systems.* The basic CDML scenario can be extended by more security requirements. For instance, the CDML system must be robust against data poisoning and model poisoning attacks, where malicious parties manipulate their local training data to compromise predictions of ML models. Moreover, in addition to training data confidentiality, parties can require the CDML system to keep ML model architectures and ML tasks confidential.

### 2.2 ML Paradigms and the CDML Scenario

This section elucidates why CML and DML are not ideally suitable for the above basic CDML scenario and introduces CDML as a more suitable ML paradigm. In the following, we assume that parties control software agents (referred to as agents) that can interact with agents from other parties.

*Collaborative Machine Learning.* CML [e.g., 64, 94, 96] comprises approaches that enable agents to develop ML models together based on their individual training data. An intuitive approach to CML is to collect training data of agents on a central server that trains an ML model. To preserve training data confidentiality in this approach, techniques like homomorphic encryption [94, 96] and trusted execution environments [66, 116] can be used. Similarly, multi-party computation allows agents to transmit secret shares of training data to a set of servers, which then execute a protocol to train an ML model without learning about the training data of each agent [64, 65]. Such CML approaches are unsuitable for the basic CDML scenario described in Section 2.1 because they require transmitting training data to central servers, either in plain or encrypted form.

*Distributed Machine Learning.* DML [e.g., 51, 60, 72, 74] was developed to accelerate training large ML models, such as deep learning models, with large amounts of training data, by distributing training tasks to multiple agents that train (parts of) ML models. In DML systems (e.g., computing clusters), ML models are trained in two ways [19, 75, 91]: data parallel and model parallel. In data parallel training, partitions of a known entire training data set are passed to agents. Each agent trains an identical ML model on different subsets of the entire training data set. In model parallel training, each agent uses identical training data but only trains a part of an ML model. In preparation for DML, training data is usually gathered by a central server, called parameter server that coordinates the DML system [19, 54]. After the local training of an ML model, agents transmit their ML model updates to the parameter server. The parameter server stores ML model updates and offers the latest parameters to agents. Agents fetch the parameters to proceed with the local training of the ML model. The central party then distributes gathered training data across agents to achieve a uniform workload for each (i.e., through identical distribution). Through the uniform workload distribution, idle times of agents can be minimized so that ML models are trained with high computational efficiency [91].

While DML can allow for collaboration between agents [74], DML approaches require transmission of (parts of) training data to a central agent. To train ML models, such a central agent gathers and reads training data and distributes it to other agents, typically in an identically distributed manner. This violates the requirement in the basic CDML scenario, described in Section 2.1, that prohibits transmitting training data between agents. Consequently, common DML approaches are unsuitable for the basic CDML scenario.

*Collaborative Distributed Machine Learning.* A combination of CML (e.g., leveraging training data from multiple agents) and DML (e.g., distributed training of ML models) forms the foundation of CDML. In CDML systems, agents receive ML tasks from other agents and use local training data and self-controlled computing resources to accomplish such ML tasks. ML tasks specify the objectives pursued with ML models (e.g., next-word prediction) and include information about the approach (e.g., the ML model architecture to be used). CDML leverages DML techniques to speed up the training process through parallelization. However, identical training data distribution, like in DML, is hard to achieve because the complete training data sets are unknown to agents in CDML systems. Thus, the performance benefits targeted in DML may not be fully leveraged [121].

CDML allows agents to retain their training data as it is not disclosed to other agents. Consequently, CDML suits the basic CDML scenario described in Section 2.1.

## 2.3 Trends in CDML

In 2013, *gossip learning* [68], one of the first CDML concepts, enabled distributed computing on training data sets that are not transferred between agents. In gossip learning, agents collaborate peer-to-peer. Each time an agent receives an ML model, it updates the ML model with its training data before passing it to other agents. Gossip learning supports training data confidentiality, low idle time of agents [6], and high availability but often entails slow convergence of ML models.

To tackle weaknesses of gossip learning, *federated learning* was introduced in 2015 [50]. In basic federated learning, a central agent orchestrates the learning process [8, 35, 63]. Soon after its invention, various shortcomings of the federated learning concept were uncovered. For example, federated learning systems were shown to be inefficient due to high communication costs [45] and prone to performance bottlenecks caused by the use of a central parameter server [117]. From a security perspective, federated learning systems bear risks of failures caused by an adversarial central parameter server [45].

To address drawbacks of federated learning, practice and research developed other CDML concepts that help ensure confidentiality of training data, ML models, and ML tasks and reduce the risks of failures caused by an adversarial central parameter server.

In *split learning* [32, 90], agents only train parts of ML models defined by a cut layer, which indicates the layers of a neural network where the complete neural network is split. Agents only receive the specifications of the cut layer as an interface to input parameters for the training of the rest of the ML model. By only disclosing parts of ML models, split learning helps to keep (at least parts of) ML models confidential. However, this gain in ML model confidentiality comes at the cost of the learning performance of ML models compared to federated learning [87].

In *swarm learning* [97], all agents use the same ML model for the training. Other than federated learning, swarm learning relies on redundant agents orchestrating the training process in peer-to-peer networks [97]. The redundant execution of tasks can make swarm learning systems more robust than federated learning systems [97]. However, extensive redundancies render swarm learning systems usually less resource-efficient and more complex than federated learning systems [108].

Despite the strong focus on training data confidentiality in CDML, training data can be reconstructed from interim results, such as gradients [7, 28, 77]. To meet industry requirements regarding data confidentiality, researchers investigated the integration of CDML with privacy-enhancing technologies (PETs), such as differential privacy [2, 99, 109] and homomorphic encryption [2, 29].

Transparency about the origin of training data is critical for successfully using CDML systems in industry, especially due to laws that impose data governance obligations across value chains in the development of ML models. To support such transparency, researchers combined CDML systems with distributed ledger technology, including blockchain technology [82, 101] The variety of combinations between CDML concepts and other technologies makes it challenging to understand and compare principal CDML concepts. A thorough analysis of the CDML paradigm is needed.

## 2.4 Related Research

Since the variety of CDML system designs increased, extant research has started to compare CDML systems to understand their commonalities and differences. Such comparisons are often based on benchmarks, for example, between federated learning, split learning, and SplitFed learning systems [87]. CDML system benchmarks help better understand likely behaviors and key traits of CDML systems, especially in terms of performance characteristics like convergence speed of ML models [87], communication cost [87], and prediction performance [113]. However, benchmark studies only shed light on a few CDML system designs, neglecting others. Moreover, benchmark studies seldom explain why CDML system designs lead to different key traits.

Studies on differences between CDML system designs revealed several design options for federated learning systems. Such design differences mainly relate to network topologies for communication (e.g., via central servers and peer-to-peer) and computational schedules [78], such as sequential training of ML models and parallel training coordinated by central servers. Key traits that originate from the different design options are discussed with a focus on confidentiality. Design differences between other CDML systems (e.g., assisted learning and split learning systems) remain unknown. In a comparison between federated learning systems, split learning systems, and SplitFed learning systems [87], key traits of those CDML systems are pointed out, with a focus on learning performance, resource consumption, ML model confidentiality, and training data confidentiality. Despite those valuable insights, several design options (e.g., regarding the network topology and computational schedules) and their influences on key traits of CDML systems remain unclear.

Since extant comparisons focus only on a few CDML concepts, it is hard to oversee the entirety of available CDML system designs. To help developers design CDML systems that meet use case requirements, CDML systems need to be comparable regarding the various CDML concepts, CDML system designs, and key traits of CDML systems.

## 3 Methods

We applied a four-step research approach to develop a CDML system conceptualization in response to RQ1 (*What are commonalities and differences between CDML systems?*) and to extract key traits of CDML systems originating from different designs to answer RQ2 (*What are the key traits of principal CDML system designs?*). First, we conceptualized CDML systems described in literature (see Section 3.1) using agent-oriented modeling [102]. Second, we used the CDML system conceptualization to develop CDML archetypes (see Section 3.2.1). Third, we extracted traits of CDML system designs from literature. We assigned the CDML system designs, including their traits, to CDML archetypes and aggregated the traits to key traits (see Section 3.2.2). Fourth, we conducted ten semi-structured expert interviews with practitioners who set up and used CDML systems in different industries and identified CDML use cases (see Section 3.2.3). The following describes the methods applied in more detail.

### 3.1 Development of the CDML System Conceptualization

To develop the CDML system conceptualization, we adopted the Gaia methodology for agent-oriented modeling [102] and analyzed CDML systems presented in literature using open coding, axial coding, and selective coding [13] as described in Section 3.1.2. The literature review revealed commonalities and differences between CDML system designs. Based on the differences between CDML system designs, we developed design options for customization of CDML systems (e.g., agent role distributions, optional communication paths, and structures of training processes). We tested and refined the CDML system conceptualization in three iterations (see Section 3.1.3).

*3.1.1 The Gaia Methodology.* The Gaia methodology supports development of agent-based models that can serve as blueprints for the implementation of software systems [102]. The Gaia methodology comprises a roles model and an interaction model that form the concept level of system descriptions. The roles model describes the tasks and basic processes, including the resources that agents can use and functions that an agent performs. Each role comprises responsibilities, permissions, activities, and protocols. Responsibilities define the functions an agent of a specific role performs. Permissions describe which resources are available to agents with specific roles. Activities are computations that agents perform locally without interaction with other agents. Protocols as part of the roles model reference protocol definitions in the interaction model. The interaction model describes how interactions between agents of specific roles are designed.

The interaction model specifies how agents with specific roles interact with each other in a purposeful way. Frequently recurring interactions between agents with other agents, objects, or the environment of the MAS are recorded as interaction patterns. Each interaction pattern is described in a protocol definition. Interactions originate from an agent (i.e., an initiator) and are directed to an interaction partner (i.e., a responder). For each interaction, the initiator prepares an input and issues the input into the interaction process. The output comprises the information the responder receives at the end of the interaction.

Building on the concept level, the Gaia methodology outlines a design level comprising a service model, an acquaintance model, and an agent model. The service model describes the main services that are necessary to implement an agent role. A service refers to a single, coherent block of activity in which an agent engages. Every activity corresponds to a service, but not every service will correspond to an activity. The services that an agent offers depend on its roles and corresponding activities and protocols.

The acquaintance model describes communication paths between different agent types in a CDML system. The acquaintance model helps to identify communication bottlenecks that may arise during run-time. The agent model describes the agent types utilized by CDML systems. Agent types are combinations of roles. Moreover, the agent model describes instances of these agent types that will populate the CDML system.

Similar to the structure of the Gaia methodology, the CDML system conceptualization describes the principal functioning of CDML systems based on roles and interaction models and design options for customizing CDML systems in the service model, acquaintance model, and agent model. The principal functioning of CDML systems showcases commonalities between CDML systems. The design options can be used to customize CDML systems.

*3.1.2 CDML System Conceptualization.* We conceptualized CDML systems in three steps: *start set compilation*, *development of an initial version of the CDML system conceptualization*, and *test and iterative refinement*. The following describes the three steps in more detail.

*Start Set Compilation.* Before starting the literature search, we specified the following inclusion criteria to systematically assess the relevance of publications on CDML (see Table 1): *English language*, *level of detail*, *topic fit*, and *uniqueness*. Only publications that meet all inclusion criteria were considered relevant for the CDML system conceptualization.

Each author independently generated a set of publications potentially relevant to developing the CDML system conceptualization. We searched for publications that cover a large variety of CDML systems and offer detailed descriptions of CDML system designs. Then, we consolidated the independently generated sets of publications into a preliminary set of relevant literature comprised of 30 publications, including peer-reviewed scientific articles and grey literature. Next, we applied the inclusion criteria to the publications in the preliminary set of relevant literature (see Table 1). We removed one publication from the preliminary set of relevant literature because it was a duplicate. We independently rated the relevance of the remaining 29 publications for the conceptualization as 'relevant,' 'maybe relevant,' and 'irrelevant' based on the inclusion criteria. Whenever we were at variance regarding the relevance of publications, we discussed until we concluded with unanimous decisions. This relevance assessment led us to exclude 18 additional publications. The final set of relevant literature comprised eleven publications.

*Development of an Initial Version of the CDML System Conceptualization.* We analyzed the publications in the set of relevant literature by applying open, axial, and selective coding [13]. In open coding, we extracted aspects of CDML systems relevant to explain their designs and functioning. We iteratively refined the coding to achieve mutual exclusiveness between the codes and exhaustiveness of the coding. For example, we merged the codes 'client' and 'device' into the code 'trainer' and the codes 'sendParameters' and 'sendGradients' into the code 'transmitInterimResult'.

In axial coding, we extracted relationships between the codes developed in open coding. For example, we identified that the code 'transmitInterimResult' can be implemented differently. We coded each implementation (e.g., 'activations' and 'gradients') and noted the relationship between 'transmitInterimResult' and 'gradients'. We classified the extracted codes into coding schemes. The coding schemes correspond to five agent-oriented models (i.e., the roles model, the interaction model, the agent model, the service model, and the acquaintance model) introduced in the Gaia methodology [102]. For example, we classified the code 'trainer' as a role in the roles model and the code 'transmitInterimResult' as a protocol in the interaction model.

Table 1. Criteria to be met for the inclusion of publications on CDML systems in the literature analysis

| Name | Descriptions |
| --- | --- |
| English Language | The publication must be in English. |
| Level of Detail | The publication must present enough details to understand the design and functioning of CDML systems. |
| Topic Fit | The publication must describe at least one CDML system design. |
| Uniqueness | The publication must not be included in the set of relevant literature. |

In selective coding, we assigned the coding results from axial coding to the overarching concept of CDML systems.

*3.1.3 Test and Iterative Refinement.* We gathered evidence for the external validity of the CDML system conceptualization by testing whether CDML systems, which we did not use to develop the conceptualization, can be successfully represented using the conceptualization.

To find additional publications that present CDML systems, we applied a backward search and a forward search [100] to the analyzed publications. We decided on the relevance of the additional publications using the previously used inclusion criteria (see Table 1). We only added publications that meet the inclusion criteria to the set of relevant literature.

We again applied open, axial, and selective coding [13] to analyze the additional publications. Based on the coding, we modeled the CDML systems presented in the additional publications using the preliminary CDML system conceptualization. When we recognized that a CDML system could not be modeled with the CDML system conceptualization, we refined the conceptualization accordingly. Subsequently, we continued with the validity test until we analyzed all relevant publications on CDML that we identified in the last round of backward and forward searches. When the CDML conceptualization needed refinement, we again started the third step ('Test and Refinement') after all publications of the respective round were analyzed. We repeated this step three times until the validity test did not uncover additional needs for refinements (see Table 2).

Table 2. Overview of the development of the CDML system conceptualization

| Category | Initial Version | Iteration 1 | Iteration 2 | Iteration 3 | Summary |
|---|---|---|---|---|---|
| Number of Publications for the Validity Test | 10 | 5 (fs), 4 (bs) | 8 (fs), 1 (bs) | 12 (fs), 4 (bs) | 44 |
| Number of CDML Systems for the Validity Test | 13 | 5 (fs), 6 (bs) | 9 (fs), 1 (bs) | 12 (fs), 4 (bs) | 50 |
| CDML Systems Successfully Modeled | n.a. | 3 | 3 | 16 | 22 |
| CDML Systems Leading to Conceptualization Refinements | n.a. | 8 | 7 | 0 | 15 |

*bs: backward search    fs: forward search    n.a.: not applicable*

## 3.2 CDML Archetype Development

Using the CDML system conceptualization, we systematically compared CDML systems to develop CDML archetypes. Moreover, we identified key traits of each CDML archetype and suitable use cases.

*3.2.1 Archetype Development.* For each CDML system analyzed in the previous step, we developed an agent model, a service model, and an acquaintance model using the CDML system conceptualization. Based on those models, we compared CDML system designs to identify systematic differences and similarities that were a foundation for developing CDML archetypes.

*Agent Model.* We started the analysis by examining role distributions in CDML systems to extract common agent types. To identify agent types and their distribution, we analyzed the agent models of the 50 CDML systems, which we previously used for testing the validity of the CDML system conceptualization (see Section 3.1.2). We developed one agent model for each of the analyzed CDML systems. Next, we compared the agent models with each other to identify similarities and differences. Based on similarities between the agent models, we classified the 50 CDML systems into 18 groups of CDML systems. Each CDML system was assigned to exactly one group.

*Service Model.* We analyzed the grouped CDML systems to identify similarities in the design options implemented for activities and protocols. If CDML systems associated with different groups showed similar uses of design options, we merged those groups into candidate CDML archetypes. For example, we merged assisted learning systems with split learning systems because both systems use the design option 'train a part of the ML model' for the protocol 'trainMLModel'. Overall, we developed six candidate CDML archetypes from the 18 groups of CDML systems.

*Acquaintance Model and Main Processes.* We analyzed the communication paths of the individual CDML systems using their acquaintance models. We merged candidate CDML archetypes whenever we observed similarities in acquaintance models of CDML systems of different archetypes. Overall, we revealed four final CDML archetypes.

*3.2.2 Identification of Key Traits of CDML Archetypes.* Using the set of relevant literature on CDML systems (see Section 3.1.2), we performed open coding [13] to extract preliminary traits of CDML systems that authors point out to highlight strengths and weaknesses of CDML system designs. We noted the referenced CDML systems for all preliminary traits and noted explanations of how the trait originates from the CDML design in axial coding [13]. For example, the key trait 'communication bottleneck' is referenced in several publications about federated learning systems. This trait originates from the reliance of federated learning systems on a central agent [92, 113, 118]. We added a description of whether the referenced CDML system has a strength or weakness in the respective trait. The analysis revealed 132 codes representing preliminary traits of 50 CDML systems. Subsequently, we harmonized the preliminary traits in three iterations to ensure mutual exclusiveness and exhaustiveness of the coding [13]. For example, we aggregated the preliminary traits 'does not rely on an orchestrator' and 'no need to rely on a third party' to the trait 'fault-tolerant.' The analysis revealed 38 traits of CDML systems.

Next, we mapped the 38 traits of CDML systems to CDML archetypes. We discussed which traits apply to all CDML systems assigned to CDML archetypes. When we reached unanimous agreement, we assigned the features that all CDML systems associated with the same CDML archetype have in common to the CDML archetype as key features. We repeated this process for all traits extracted from the literature analysis.

Table 3. Demographic data of the interviewees

| Years of Work Experience | Job Position | Highest Degree of Education |
| --- | --- | --- |
| < 5 (3) | AI Consultant (1), Business Developer (2), | Bachelors (2) |
| 5 - 10 (4) | CEO (3), Co-founder (2) | Masters (5) |
| 20 - 25 (2) | Data Scientist (1), Research Engineer (1) | Masters equivalent (1) |
| 25 - 30 (1) | | PhD (2) |

*3.2.3 Identification of Use Cases per Archetype.* We conducted semi-structured interviews with CDML experts (see Table 3) to uncover use cases for the identified CDML archetypes. In preparation for the interviews, we developed an interview guide [80] that comprised four sections. Section 1 (*introduction*) explained the objectives of the interviews and introduced the foundations of CDML. In Section 2 (*use case characteristics*), the interviewees were asked to describe CDML use cases they worked on, including questions about the used CDML system design and the reasons for using that design. Section 3 (*demographics*) included questions about the interviewee's demographics. In Section 4 (*conclusion*), we asked interviewees for their final thoughts on the topic. We tested the interview guide in a pilot interview and revised the guide according to the feedback we obtained.

Based on personal contacts and cold acquisition via email and social media, we reached out to CDML experts from companies offering services to operate CDML systems. We invited CDML experts from different fields with experience in setting up, operating, and using CDML systems to obtain a broad overview of CDML use cases.

We conducted ten one-to-one interviews with experts from nine companies. The interviews took, on average, 53 minutes and, in sum, 535 minutes. All interviews were recorded, pseudonymized, and transcribed verbatim.

We extracted CDML use cases from the transcribed interviews. For each use case, we defined its main purpose, collaborators, type of collaboration, and the CDML system used. For example, we extracted the purpose *organ rejection detection support*, the collaborators *university hospitals*, the type of collaboration *inter-organizational*, and the CDML system *federated learning*.

## 4 A CDML System Conceptualization

This section introduces a CDML system conceptualization, describing the principal functioning and design options for the customization of CDML systems. The principal functioning describes the foundational design of CDML systems and is represented in a roles model and an interaction model (see Section 4.1). The design options are summarized in a service model, an acquaintance model, and an agent model (see Section 4.2). Design options refer to alternatives for implementing specific aspects of the CDML systems, such as the assignment of agent types (i.e., combinations of roles) to agents in CDML systems and the definition of types of interim results to be transmitted between agents (e.g., neural network gradients). Agents in specific roles execute activities and protocols associated with their assigned roles. Agents keep their roles until the coalition dissolves. Agents can activate or deactivate roles depending on the tasks to work on.

### 4.1 Principal Functioning of CDML Systems

Every CDML system passes through three phases of the CDML life cycle: initialization phase, operation phase, and dissolution phase. In the initialization phase, agents form and initialize a coalition that can become a CDML system. In the operation phase, agents interact to train or use ML models. In the dissolution phase, agents terminate their collaboration and dissolve the CDML system. The subsequent paragraphs describe each phase of the CDML life cycle with a focus on the roles (see Table 4) and interactions between roles in each phase.

Table 4. Overview of roles and corresponding activities and protocols in the CDML conceptualization

| Role | Description | Activities | Protocols |
| --- | --- | --- | --- |
| Configurator | Approves agent applications for the coalition and defines the ML model and interim results to be transmitted in the coalition | `defineInitialMLModel`, `defineInterimResult`, `registerCoalition`, `awaitApplications`, `decideOnApplication` | `applyForCoalition`, `informApplicant`, `provideMLTask` |
| Coordinator | Approves agent applications for the coalition and assigns communication paths among agents | `awaitApplications`, `decideOnApplication` | `applyForCoalition`, `informApplicant`, `provideMLTask`, `assignInterimResultRecipient` |
| Selector | Chooses agents to train and update ML models | `awaitReadinessSignal`, `selectAgent` | `applyForCoalition`, `announceAgentSelection` |
| Trainer | Uses local data and computing resources under its control to update parameters of (parts of) ML models (e.g., to produce interim results) and transmits them to other agents in the coalition | `awaitSelectionSignal`, `trainMLModel` | `applyForCoalition`, `signalReadiness`, `transmitInterimResult` |
| Updater | Uses interim results received from other agents to compute updated parameters of (parts of) its ML model | `awaitSelectionSignal`, `awaitInterimResults`, `updateMLModel` | `applyForCoalition`, `signalReadiness`, `transmitInterimResult` |

*Initialization Phase.* In the initialization phase, agents form a coalition of at least two agents to accomplish an ML task. Coalition forming is triggered by a *configurator* agent. The *configurator* agent stores the CDML system specifications about the purpose of envisioned CDML systems (i.e., the general prediction problem that ought to be addressed) and requirements for agents that are appreciated to join the coalition (e.g., in terms of the training data structure needed). As a part of the CDML system specifications the *configurator* agent defines (parts of) the initial ML model (activity: `defineInitialMLModel`) to be trained. Definitions of the (parts of) initial ML models are, for instance, the (first) layers of neural networks, a (sub-) set of parameters of linear regressions, activation functions, and the ML model architecture. Moreover, the *configurator* agent defines the structure and type of interim results (activity: `defineInterimResult`) to be transmitted between agents in the envisioned CDML system. Interim results are updates that agents compute based on local training data and the locally available (part of an) ML model. Then, the *configurator* agent registers the coalition (activity: `registerCoalition`) with a repository and starts an application process.

Agents fetch the CDML system specifications from the repository. Based on the CDML system specifications, agents decide whether to participate in the CDML system. Agents that decide to participate submit an application, including the roles they apply for, to the *configurator* agent (protocol: `applyForCoalition`). Commonly, agents can apply for the roles *coordinator*, *selector*, *trainer*, and *updater*. The *configurator* agent iteratively checks for applications from agents (activity: `awaitApplications`). Upon application receipt, the *configurator* agent decides whether to accept or reject the agent for the CDML system (activity: `decideOnApplication`). Then, the *configurator* agent responds to the applying agent with an acceptance message or a rejection message (protocol: `informApplicant`).

The *coordinator* agent assigns *trainer* agents to *updater* agents, who will be the recipients of interim results in the operation phase. Then the *coordinator* agent informs the respective agents about the assignment (protocol: `assignInterimResultRecipient`).

The *configurator* agent sends the ML task to agents in the coalition (protocol: `provideMLTask`). ML tasks are a collection of information required to train and update ML models and can include the initial ML model definition and the interim result definition.

At the end of the initialization phase, the coalition comprises at least two agents that, in combination, have the following roles: *configurator, coordinator, selector, trainer*, and *updater*. Agents may have multiple roles. We describe common combinations of roles in the design options for customizing CDML systems (see Section 4.2).

*Operation Phase.* During the operation phase, the *coordinator* agent handles applications of agents on behalf of the *configurator* agent and executes the activities `awaitApplications`, `decideOnApplication` and the protocols `informApplicant` and `provideMLTask`.

In the operation phase, agents participate in the training and execution of ML models according to their assigned roles. At the beginning, the *trainer* agent and the *updater* agent signal their readiness to the *selector* agent (protocol: `signalReadiness`). Agents that have signaled their readiness iteratively check for triggers from the *selector* agent to execute activities and protocols required to collaboratively train and update ML models (activity: `awaitSelectionSignal`).

The *selector* agent waits until agents signal their readiness (activity: `awaitReadinessSignal`) and then selects *trainer* agents and *updater* agents (activity: `selectAgent`) to activate at least one of these roles. After selection, the *selector* agent requests the selected agents to act in the corresponding roles (protocol: `announceAgentSelection`). Agents that are selected to activate the role *trainer* use their locally available (parts of an) ML model and local training data to compute interim results (activity: `trainMLModel`). *Trainer* agents send their interim results to their assigned *updater* agent (protocol: `transmitInterimResult`). The *updater* agent waits until it receives interim results (activity: `awaitInterimResults`) and uses the interim results received from *trainer* agents to compute a new version of the locally available (part of the) ML model (activity: `updateMLModel`). The execution order of training, updating, and transmitting interim results can vary between CDML systems (see Section 4.2). The procedure outlined in the operation phase is typically executed repeatedly. Protocols and activities may be executed in parallel or sequentially.

*Dissolution Phase.* In the dissolution phase, agents stop executing the processes of the operation phase. This can be the case if agents decide that (parts of) the ML model(s) have been sufficiently trained or if agents do not use the ML model anymore. When agents end their collaboration, the CDML system dissolves.

### 4.2 Design Options for the Customization of CDML Systems

We structured the identified design options, which can be used to customize CDML systems, according to the three models described in the Gaia methodology in Section 3.1.1: service model, acquaintance model, and agent model. The following presents those three models.

*4.2.1 Service Model.* The principal functioning of CDML systems can be customized in terms of activities and protocols (see Table 4) using different design options. Because service models are usually highly context-dependent [102], the following descriptions of design options for the key activities and protocols are intended as a foundation for developing detailed service models.

*Activities.* We identified 12 design options for the activities awaitApplications, selectAgent, awaitInterimResults, updateMLModel, and trainMLModel.

awaitApplications: The activity awaitApplications describes the process of iteratively checking for applications from agents that want to join the coalition. In some CDML systems, agents must communicate during the initialization phase to effectively prepare for the subsequent operation phase (e.g., in split learning systems, agents must coordinate how the ML model is split among agents). Conversely, other CDML systems (e.g., federated learning and swarm learning) necessitate less organizational effort from agents during the initialization phase, enabling agents to join after the initialization phase and still successfully participate in the operation phase. This leads to two design options for the activity awaitApplications that describe at which point agents can successfully apply to be part of the coalition. First, the agent population awaits agent applications to join the coalition 'only during the initialization phase'. Applications are ignored when the CDML system has already been initialized. For example, in most variants of split learning systems [90], the ML model layers to be trained need to be assigned to agents during the initialization phase, which prevents agents from joining after the initialization phase. Second, the agent population accepts applications 'always' [97]. This allows agents to join the CDML system at any time.

selectAgent: The activity selectAgent describes the process of deciding which agents should execute which activities and protocols. We identified three different design options that selector agents can choose from to execute the activity selectAgent. First, agents can be selected for a role 'based on votes from other agents' in the CDML system. The *selector* collects the votes of other agents and decides which agents should execute which activities and protocols; for example, all agents in the CDML system can vote on which agent activates the *updater* role and executes the updating of the ML model (activity: updateMLModel) [97]. Second, agents can be selected 'based on agent attributes', for example, based on the size of agents' datasets [31]. Third, agents can be selected 'randomly' to activate a role and execute corresponding activities and protocols [81, 107].

awaitInterimResults: The activity awaitInterimResults describes the process of iteratively checking for interim results receipt. Ensuring liveness in CDML systems requires careful consideration of the waiting time for updater agents to receive interim results from other updater agents or trainer agents. We identified two design options for implementing this activity. First, the *updater* agent waits for a specified number of interim results before updating the ML model with the interim results received, which would be 'response-bound' [84]. 'Response-bound' waiting for interim results can decrease the liveness of CDML systems if set too high; for example, when an *updater* agent awaits interim results from all *trainer* agent, but one *trainer* agent may have crashed, the *updater* agent may theoretically wait infinitely. The second design option, 'time-bound' waiting, tackles this issue [8]. If the waiting time exceeds a specified time bound, the *updater* agent updates the ML model with all interim results received during the waiting period. However, 'time-bound' waiting may lead the *updater* agent to ignore interim results received too late.

updateMLModel: The activity updateMLModel describes the process of computing a new version of the locally available (part of an) ML model using interim results from other agents. The updateMLModel activity, executed by *updater* agents, involves the enhancement of an ML model by aggregating two or more interim results received from other agents. The method of aggregating these interim results can vary and has an impact on various metrics of the ML model, including generalization error, robustness, and fairness. We identified two design options for implementing the updateMLModel activity. First, *updater* agents can perform 'batched updates' [31, 56, 61, 92]. In 'batched updates', *updater* agents use a set of interim results received from *trainer* agents to update their ML model. Second, *updater* agents can perform 'individual updates' to separately update the ML model for each interim result received from a *trainer* agent or an *updater* agent [84, 90].

trainMLModel: The activity trainMLModel describes the process of using the locally available (part of the) ML model and local training data to compute interim results. Trainer agents locally train their available (part of an) ML model. We identified three design options to customize the trainMLModel activity. First, *trainer* agents can 'train two complete ML models'. One local ML model learns representations of the training data, and a global ML model is trained on the local ML model instead of the raw training data. An advantage of this approach is that the local ML model can protect confidential attributes from the global ML model. Moreover, communication efficiency can be improved because the global ML model requires fewer parameters [46, 58]. Second, *trainer* agents can 'train one complete ML model'. A complete ML model refers to the entire set of parameters comprising the ML model. In most CDML systems, *trainer* agents store and train one complete ML model [35, 62]. Third, *trainer* agents can 'train a part of an ML model'. A part of an ML model refers to a subset of ML model parameters, such as subsets of neural network layers or linear regression coefficients. Training only a part of an ML model has two main advantages: (1) *Trainer* agents require less storage and computing resources, and (2) confidentiality of the complete ML model is better protected because *trainer* agents have access to a part of the ML model [90, 105].

*Protocols.* We identified nine design options to customize the protocols provideMLTask, announceAgentSelection, and transmitInterimResult.

provideMLTask: The protocol provideMLTask describes the process of transmitting data required to train and update ML models. A pivotal aspect in designing CDML systems involves determining the extent of information the *configurator* imparts to other agents in the initialization phase. This decision is crucial because the level of detail about the ML task shared by the *configurator* directly correlates with disclosing the *configurator* goals, rendering the configurator susceptible to confidentiality and robustness threats. We identified two design options for the protocol provideMLTask. First, the *configurator* can 'provide only interim result definition' to other agents in the CDML system. The *configurator* agent only provides the interface between agents (e.g., exchange parameters or gradients). The exact ML model to be used remains unknown to other agents, encompassing the ML model architecture and its hyperparameters [105]. Second, the *configurator* agent 'provides ML model definition and interim result definition', which discloses the goals of the *configurator* [e.g., 8, 63].

announceAgentSelection: The protocol announceAgentSelection refers to the communication between agents that triggers the activation of roles. In some CDML systems [e.g., in assisted learning systems; 16, 105], *trainer* agents must possess identical training sample IDs within their respective local training datasets. In such CDML systems, after selecting *trainer* agents for a training round, it becomes necessary to communicate which specific training samples they are assigned for training. In other CDML systems [e.g., in swarm learning systems; 97], such communication is unnecessary or impractical, because agents have different training sample IDs. We identified two design options for the protocol `announceAgentSelection`. First, when using the design option 'only role', the *selector* agent announces which agent should activate what role [8, 39, 97]. Second, when using the design option 'role and training sample IDs', the *selector* agent announces which agents should activate which role and the training sample IDs that should be trained [16, 105].

`transmitInterimResult`: The protocol `transmitInterimResult` describes the process of sending interim results and labels, if necessary to use the interim results, to other agents. Different definitions of interim results lead to five design options for the `transmitInterimResult` protocol. First, agents can transmit 'parameter values' [53, 104]. Parameter values refer to a set of variables or weights that an ML model learns from training data and determine how the ML model makes predictions based on input data. Second, agents can transmit 'gradients' [63, 84]. Gradients refer to the partial derivatives of a loss function with respect to model parameters and indicate the direction and magnitude of adjustments needed to reduce prediction error. Third, agents can transmit 'activations with labels' [32, 90]. We refer to activations as intermediate outputs of an ML model for a given input. When the ML model is presented with input data, it propagates the data through its layers, applies the learned parameters (weights and biases), and produces an output. We refer to the output as activations if it is not the final output of the ML model. If the output is from the final layer or includes all parameter values of the ML model, we call the outputs predictions. Fourth, agents can transmit 'activations without labels' [32, 90]. Fifth, agents can transmit 'pseudo-residuals' [16]. Pseudo-residuals refer to differences between actual target values and predicted values produced by an ML model.

*4.2.2 Acquaintance Model.* CDML systems have indispensable communication paths, which are essential in each CDML system, and dispensable ones, which are part of only special CDML systems. Because communication paths differ between the life cycle phases of CDML systems, the following describes indispensable and dispensable communication paths (design options) for each phase separately.

*Initialization Phase.* *Configurator* agents must have a bidirectional communication path to all other agents for two purposes: first, to participate in the coalition application process (protocols: `applyForCoalition`, `informApplicant`); second, to provide them with the ML task definition (protocol: `provideMLTask`). *Coordinator* agents must have unidirectional communication paths to *trainer* agents to inform them about the *updater* agents to send their interim results to (protocol: `assignInterimResultRecipient`). Such communication paths enable formation of sub-coalitions around *updater* agents [8, 93, 104]. *Coordinator* agents may have unidirectional communication paths to *updater* agents. Via such communication paths, *coordinator* agents can inform *updater* agents to which *updater* agents they should send intermediate results (protocol: `assignInterimResultRecipient`). Such communication paths can be used for a hierarchically organized CDML system, in which *updater* agents communicate with each other to improve their local ML model without using local training data [8, 93, 104].

*Operation Phase.* *Selector* agents must have bidirectional communication paths to *trainer* agents and *updater* agents. Such communication paths enable *selector* agents to receive signals that these agents are ready to participate in the training (protocol: `signalReadiness`) and to inform these agents that they are selected for the training (protocol: `announceAgentSelection`). *Trainer* agents must have unidirectional communication paths to *updater* agents to send interim results (protocol: `transmitInterimResult`). *Coordinator* agents can have bidirectional communication paths to all other agent roles if applications can be received and processed after the initialization phase. In this case, *coordinator* agents take over handling applications from *configurator* agents (protocols: `applyForCoalition`, `informApplicant`). Because agents can apply and be admitted to a CDML system in the operation phase, this communication path enables the CDML system to address issues in the agent population. For example, if training data is insufficient in the operation phase, more *trainer* agents can be admitted to the CDML system. *Updater* agents can have unidirectional or bidirectional communication paths with another *updater* agent to exchange information about their ML model update [8, 104]. Such communication paths allow for hierarchical structures with more than one *updater* agent. *Trainer* agents can have bidirectional communication paths to *updater* agents, for example, to send and receive interim results (protocol: transmitInterimResult). Such bidirectional communication paths are common in CDML systems. In some CDML systems [e.g., one-shot federated learning; 31], *trainer* agents send interim results to *updater* agents without receiving interim results [31].

*Dissolution Phase.* The communication paths between agents are dissolved in the dissolution phase. Agents that store a local ML model can keep and use it for their own predictions.

Table 5. Overview of common agent types in CDML systems

| | | **Agent Roles** | | | | |
|---|---|---|---|---|---|---|
| | | Configurator | Coordinator | Selector | Trainer | Updater |
| **Agent Type** | Tra | | | | ✓ | |
| | CooSel | | ✓ | ✓ | | |
| | TraUpd | | | | ✓ | ✓ |
| | ConTraUpd | ✓ | | | ✓ | ✓ |
| | ConCooSelUpd | ✓ | ✓ | ✓ | | ✓ |
| | CooSelTraUpd | | ✓ | ✓ | ✓ | ✓ |
| | ConCooSelTraUpd | ✓ | ✓ | ✓ | ✓ | ✓ |

*4.2.3 Agent Model.* Agent types are combinations of roles presented in the roles model (see Table 4). CDML systems require at least two agents that, in combination, have the following roles (see Section 4.1): *configurator*, *coordinator*, *selector*, *trainer*, and *updater*. These roles are commonly assigned to agents in seven combinations (see Table 5), each combination forming one agent type. Identical agent types can be assigned to multiple agents, for example, to increase redundancies in the processing of ML tasks [97] or to distribute workload in the processing of ML tasks [8]. We constructed the name of each agent type by concatenating the first three letters of the combined roles in alphabetical order. For example, an agent type that comprises the roles *configurator*, *trainer*, and *updater* becomes a *ConTraUpd* agent. We describe the seven agent types in the following and exemplify occurrences of each agent type in different CDML systems.

*Tra Agent Type.* The *Tra* agent type only comprises the role *trainer*. In some CDML systems, such as one-shot federated learning [31], *trainer* agents only send interim results to *updater* agents without receiving updated interim results back.

*CooSel Agent Type.* The *CooSel* agent type comprises the roles *coordinator* and *selector*. In CDML systems based on peer-to-peer networks [e.g., 38, 39], each *trainer* agent has a bidirectional communication path to all *updater* agents in the system to transmit interim results. To define which *trainer* agents exchange interim results, *trainer* and *updater* agents rely on *coordinator* agents to assign communication paths between agents during the initialization phase. To trigger *trainer* and *updater* agents to activate assigned roles in specific training rounds, *selector* agents are needed to select agents for specific training rounds. In CDML systems based on peer-to-peer networks, the *coordinator* role and the *selector* role are often combined in one agent type, *CooSel*, that employs sophisticated rules to assign and select agents effectively. For example, metropolized random walk with backtracking is employed to achieve unbiased agent sampling in large and dynamic unstructured peer-to-peer networks [85].

*TraUpd Agent Type.* The *TraUpd* agent type combines the roles *trainer* and *updater*. In basic federated learning systems [e.g., 62, 63, 98], *trainer* agents train their ML models on local data and send interim results to *updater* agents. The *updater* agents update the interim result with at least one other interim result and send the updated interim result back to the *trainer* agents. Agents with the role *trainer* then also act in the *updater* role by updating their ML model with received interim results. In contrast to the *Tra* agent type that is implemented in CDML systems in which the interim results are only sent unidirectional, the implementation of this agent type allows for faster convergence of ML models and is implemented in many federated learning systems.

*ConTraUpd Agent Type.* The *ConTraUpd* agent type combines the roles *configurator*, *trainer*, and *updater*. In CDML systems in which agents train different (parts of) ML models [e.g., in assisted learning systems and split learning systems; 90, 105], agents of the *TraUpd* type additionally need the ability to define their own (parts of an) ML model that they will train and update. To do so, *TraUpd* agents in such CDML systems additionally have the role *configurator* to define their own (part of an) ML model.

*ConCooSelUpd Agent Type.* The *ConCooSelUpd* agent type combines the roles *configurator*, *coordinator*, *selector*, and *updater*. In basic federated learning systems [e.g., 62, 63], there are multiple agents with the agent type *TraUpd* and one central server with the agent type *ConCooSelUpd*.

*CooSelTraUpd Agent Type.* The *CooSelTraUpd* agent type combines the roles *coordinator*, *selector*, *trainer*, and *updater*. Blockchain-based CDML systems [e.g., federated learning systems and swarm learning systems; 47, 56, 97] tend to be robust due to a high level of redundancy. To increase redundancy, almost all agents within such CDML systems have identical roles. As all agents in such CDML systems work on the same ML task, only one agent serves as the *configurator* responsible for setting up the ML task. All other agents fulfill all roles except for the *configurator* role and are thus *CooSelTraUpd* agents.

*ConCooSelTraUpd Agent Type.* The *ConCooSelTraUpd* agent type combines the roles *configurator*, *coordinator*, *selector*, *trainer*, and *updater*. Similar to *CooSelTraUpd* agents used in blockchain-based CDML systems, CDML systems based on peer-to-peer networks, such as the braintorrent system [81] and gossip learning systems [39], endow agents with a high degree of autonomy. Each *ConCooSelTraUpd* agent can initiate training of ML models.

## 5 CDML Archetypes

We developed four CDML archetypes (i.e., the confidentiality archetype, the control archetype, the flexibility archetype, and the robustness archetype) with different designs and key traits. The CDML archetypes are mainly distinguished by their agent models, acquaintance models, and principal functioning (which incorporates service models). Table 6 gives an overview of the four CDML archetypes that we describe in detail in the following.

### 5.1 Confidentiality Archetype

The confidentiality archetype is suitable for use cases in which confidentiality of ML models, ML tasks, and training data must be preserved. Agents only store parts of ML models; complete architectures of ML models trained in the confidentiality archetype are not disclosed. Instead, the global ML model is distributed across several agents, which only store parts of it. ML models are not synchronized coalition-wide during ML model training and for ML model inference. Exemplary CDML systems of the confidentiality archetype are assisted learning [10, 105], gradient assisted learning [16], FDML [41], FedLite [95], split learning [12, 32, 59, 90], SplitFed learning [43, 87], and hierarchical SplitFed learning [104].

*Agent Model.* The confidentiality archetype comprises the agent types *ConCooSelUpd* [e.g., the server in split learning; 12, 90] and *ConTraUpd* [e.g., the client in split learning; 12, 90]. In its basic configuration, the confidentiality archetype comprises one *ConCooSelUpd* agent and at least one *ConTraUpd* agent.

*Acquaintance Model.* In the confidentiality archetype, the *ConCooSelUpd* agent can communicate with all *ConTraUpd* agents on bidirectional communication paths (see Figure 1). *ConTraUpd* agents do not communicate with each other directly.

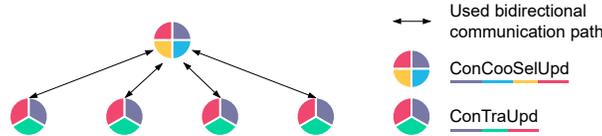

Fig. 1. Exemplary acquaintance model of the confidentiality archetype

*Principal Functioning.* In the initialization phase, a *ConCooSelUpd* agent configures its local part of the ML model and defines the interim results to be transmitted (activities: defineInitialMLModel, defineInterimResult). Local parts of the ML model can be specific layers of a neural network in split learning or just parts of a layer of a neural network in vertical split learning [90]. Examples of interim results include activations of a particular layer of a neural network [e.g., referred to as the cut layer in split learning; 90] or pseudo-residuals [16]. The *ConCooSelUpd* agent then provides *ConTraUpd* agents with the interim result definition (protocol: provideMLTask; design option: provide only interim result definition). After receiving the interim result definition, *ConTraUpd* agents individually set up their local parts of the ML model according to the interim results definition. For example, the *ConTraUpd* agents in split learning systems set up the layers of a neural network from the input layer to the cut layer. The number of outputs of the cut layer is set depending on interim result definitions.

The operation phase starts with the *ConTraUpd* agents signaling their readiness to the *ConCooSelUpd* agent (protocol: signalReadiness) to participate in the subsequent training round. Then, *ConTraUpd* agents wait for a response (activity: awaitSelectionSignal). The *ConCooSelUpd* agent decides which *ConTraUpd* agents to select for the next training round (activity: selectAgent). For example, the selection can be made based on agent attributes or randomly. After the selection, the *ConCooSelUpd* agent announces its decision to the *ConTraUpd* agents (protocol: announceAgentSelection). Selected *ConTraUpd* agents train their parts of the ML model (activity: trainMLModel; design option: train a part of the ML model) and transmit their interim results to the *ConCooSelUpd* agent (protocol: transmitInterimResult; design option: activations with labels, pseudo-residuals). The *ConCooSelUpd* agent waits for incoming interim results (protocol: awaitInterimResults). The *ConCooSelUpd* agent uses interim results to update (and train) its local (part of the) ML model (activities: trainMLModel, updateMLModel). In case of several interim results being used to update the local (part of the) ML model, the interim results can be (weighted) averaged [16, 87]. Depending on the implementation, the *ConCooSelUpd* agent transmits an interim result back to the *ConTraUpd* agents (protocol: transmitInterResult; design option: gradients). *ConTraUpd* agents use it to update their local part of the ML model. The *ConCooSelUpd* agent decides how often this process is repeated.

Table 6. Overview of CDML Archetypes and Their Designs

| | | CDML Archetypes | | | |
|---|---|---|---|---|---|
| | | Confidentiality Archetype | Control Archetype | Flexibility Archetype | Robustness Archetype |
| **Distinguishing Key Traits** | Hierarchy | Strong | Strong | Weak | Weak |
| | Coalition-wide ML Model Synchronization | No | Yes | No | Yes |
| **Design Options for Activities** | awaitApplications | Always, Only during the initialization phase | Always, Only during the initialization phase | Always, Only during the initialization phase | Always, Only during the initialization phase |
| | awaitInterimResults | Waiting for a response-threshold, Waiting for a time-threshold | Waiting for a response-threshold, Waiting for a time-threshold | Waiting for a response-threshold, Waiting for a time-threshold | Waiting for a response-threshold, Waiting for a time-threshold |
| | selectAgent | Based on agent attributes, Randomly | Based on agent attributes, Randomly | Based on agent attributes, Randomly | Based on agent votes from other agents |
| | trainMLModel | Train a part of the ML model | Train a part of the ML model, Train one complete ML model, Train two complete ML models | Train one complete ML model, Train two complete ML models | Train one complete ML model |
| | updateMLModel | Batched update, Individual update | Batched update, Individual update | Batched update, Individual update | Batched update |
| | announceAgentSelection | Role and training sample IDs | Only role | Only role | Only role |
| **Design Options for Protocols** | provideMLTask | Provide only interim result definition | Provide ML model definition and interim result definition | Provide ML model definition and interim result definition | Provide ML model definition and interim result definition |
| | transmitInterimResult | Activations with labels, Activations without labels, Gradients, Pseudo-Residuals | Gradients, Parameter values | Gradients, Parameter values | Gradients, Parameter values |
| **Agent Types (Number of Occurences)** | | ConCooSelUpd (1), ConTraUpd (1 ≤) | ConCooSelUpd (1), TraUpd (1 ≤) | ConCooSelTraUpd (1 ≤), CooSelTraUpd (1 ≤) | ConCooSelTraUpd (1), CooSelTraUpd (1 ≤) |
| **Exemplary CDML Systems** | | Assisted learning systems [10, 105], Split learning systems [32, 59, 90], SplitFed learning systems [43, 87] | Federated learning systems [8, 15, 17, 31, 58, 63, 84, 92, 118] | BrainTorrent systems [81], Decentralized federated learning systems [46, 52, 61, 106, 110, 113] | Blockchain-based federated learning systems [47, 53, 56, 107], Swarm learning systems [82, 97] |
| **Exemplary Use Case Purposes** | | Inter-organizational: Optimizing raw material processes in supply chains | Intra-organizational: Company audits, Predictive maintenance, Quality inspection Inter-organizational: Organ rejection detection, clinical trials | Intra-organizational: Spam filtering Inter-organizational: Brain segmentation | Inter-organizational: Disease diagnostic |
| **Tools** | | Simulation: SLPerf [120], PaddleFL [70], GAL [14] Deployment: PySyft [122], CoIDA [18] Combination with PETs:PySyft [122], PaddleFL [70] | Simulation: FedML [37], Flower [5], OpenFL [76], NvideaFlare [79], PaddleFL [70] Tensorflow Federated [30], FedN [21] Deployment: FedML [37], Tensorflow Federated [30], OpenFL [76], PaddleFL [70] FATE [20], Flower [5], NvideaFlare [79], FedN [21] Combination with PETs: FATE [20], NvideaFlare [79], OpenFL [76], PaddleFL [70] PySyft [122] | Simulation: GoLF [67], GossipFL [86] | Simulation: DFL [88], Deployment: Swarm learning library [40], DFL library [88] |

*5.1.1 Key Traits.* The confidentiality archetype relies on a strongly hierarchical agent organization and does not have a coalition-wide synchronization of ML models. Because ML models are not synchronized among agents, each agent stores a different (part of an) ML model, which can remain confidential. The most important trait of the confidentiality archetype is that it preserves training data confidentiality and ML model confidentiality. Moreover, the confidentiality archetype is usually efficient in computing and storage because agents only store and compute parts of ML models [90, 95]. The confidentiality archetype requires fewer training rounds than the control archetype and ML models commonly converge quickly [32, 90]. The confidentiality archetype has high communication costs incurred by ML model partitioning and the necessary communication of both activations and gradients [95].

Some CDML systems that correspond to the confidentiality archetype, such as split learning systems [90], can have high idle times of trainer agents since the trainer agents only interact with the updater agents sequentially [87]. Other CDML systems, such as SplitFed learning systems [43, 87], address this issue by combining elements of split learning and federated learning and, thus, can reduce the idle times [87]. As no agent has access to the entire ML model, the coalition (or a subset of it) is required to make ML model inferences. Therefore, the coalition can only be resolved when the ML model is not used anymore. Despite its focus on confidentiality, CDML systems that correspond to the confidentiality archetype can still be vulnerable to model inversion, model stealing, and inference attacks without additional protection measures [22, 71]. Moreover, a malicious agent may poison the ML model [26, 111].

*5.1.2 Variants of the Confidentiality Archetype.*

*U-Shaped Split Learning [90].* U-shaped split learning systems can be used to train neural networks. A selected `ConTraUpd` agent executes the forward propagation up to a specific layer (i.e., the first cut layer) and only transmits activations to the *ConCooSelUpd* agent (protocol: `transmitInterimResults`; design option: `activations without labels`). The *ConCooSelUpd* agent continues the forward propagation up to the second cut layer and transmits activations back to the *ConTraUpd* agent. The *ConTraUpd* agent completes the forward propagation, starts the backpropagation, and transmits the gradients of the second cut layer to the *ConCooSelUpd* agent (protocol: `transmitInterimResults`; design option: `gradients`). Using these gradients, the *ConCooSelUpd* agent continues the backpropagation to the first cut layer and transmits the gradients of the first cut layer to the *ConTraUpd* agent. The *ConTraUpd* agent executes the backpropagation for the remaining layers and, thus, completes a training round.

*5.1.3 Use Cases.* CDML systems of the confidentiality archetype remain rarely used in industrial use cases. Exemplary use cases are in inter-organizational settings with high confidentiality requirements. In manufacturing, for instance, split learning has been used to develop ML models that aid in *optimizing raw material processes* in a supply chain. Collaborating organizations are suppliers and consumers within the supply chain that aim to keep ML models and ML tasks confidential from competitors.

*5.1.4 Tools.* Different tools were developed to help simulate and deploy CDML systems of the confidentiality archetype. For the *simulation* of such CDML systems, tools like *SLPerf* [120], *PaddleFL* [70], and *GAL* [14] can be used. Tools that facilitate the *deployment* of CDML systems of the confidentiality archetype include *PySyft* [122], *PaddleFL* [70], and *CoIDA* [18].

## 5.2 Control Archetype

The control archetype is suitable for use cases where a single agent should coordinate the CDML system in a top-down manner. The control archetype incorporates a hierarchical communication structure with an agent on the top level that controls the training process. The agent on top receives all interim results and synchronizes the training process by deciding on the global ML model to be trained in each training round. Exemplary CDML systems of the control archetype implement variants of federated learning [8, 58, 63, 84], including HeteroFL [15], and hierarchical federated learning [92, 118], one-shot federated learning [31], and SemiFL [17].

*Agent Model.* The control archetype comprises the agent types *ConCooSelUpd* [e.g., the server in federated learning systems; 8] and *TraUpd* [e.g., clients in federated learning systems; 8]. In its basic form, the control archetype comprises one *ConCooSelUpd* agent and at least one *TraUpd* agent.

*Acquaintance Model.* The acquaintance model of the control archetype has the structure of a tree (see Figure 2). Agents bidirectionally communicate in a strictly hierarchical manner along the vertexes of the tree. In its basic form, there are two hierarchical levels [63]. A root *ConCooSelUpd* agent resides at the top level of the hierarchy. At least one *TraUpd* agent resides at the bottom level of the hierarchy. There can be additional levels between the top level and the bottom level [92, 118]. The inner nodes of the tree are *ConCooSelUpd* agents, whereas the leaves represent *TraUpd* agents.

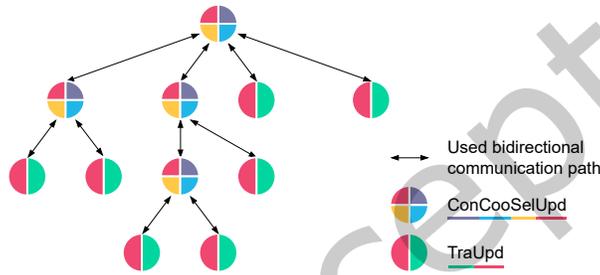

Fig. 2. Exemplary acquaintance model of the control archetype

*Principal Functioning.* In the initialization phase, the *ConCooSelUpd* agent at the top level of the hierarchy defines the initial ML model and interim results (activities: defineInitialMLModel, defineInterimResult). Suppose there are additional *ConCooSelUpd* agents on lower levels of the acquaintance model. The initial ML model and interim result definitions are propagated to those agents by executing the protocol *provideMLTask* (design option: ML model definition and interim result definition). *ConCooSelUpd* agents on lower levels of the acquaintance model can only forward parts of the ML model (i.e., sub-models) to their subordinate nodes. Thus, each *ConCooSelUpd* agent can individually define the initial ML model and interim results for its descendants (activities: defineInitialMLModel, defineInterimResult).

In the operation phase, *TraUpd* agents execute the *signalReadiness* protocol to signal their availability to participate in a training round to their respective superordinate *ConCooSelUpd* agent. Then, *TraUpd* agents wait for a selection signal (activity: awaitSelectionSignal). *ConCooSelUpd* agents decide which of their subordinate *ConCooSelUpd* and *TraUpd* agents to include in a training round. Once a sufficient number of subordinate agents have signaled their readiness to a *ConCooSelUpd* agent, it signals its readiness to its superordinate agent and waits for a selection signal (activity: awaitSelectionSignal). This process is repeated recursively throughout the hierarchy until it reaches the root *ConCooSelUpd* agent. Then, the root *ConCooSelUpd* agent selects (a subset of) its subordinate agents to participate in the upcoming training round (activity: selectAgent; design option: based on agent attributes or randomly) and announces its selection to its subordinate agents (protocol: announceAgentSelection). Afterward, the root *ConCooSelUpd* agent transmits the current version of the ML model, or a part thereof, to subordinate agents (protocol: transmitInterimResult; design option: gradients or parameter values) and waits for interim results (activity: awaitInterimResult; design option: waiting for a time-threshold or waiting for a response-threshold). This selection process is repeated by subordinate *ConCooSelUpd* agents until the leaf *TraUpd* agents.

The *TraUpd* agents update their local ML model based on the interim result received (activity: updateMLModel; design option: batched update) and train their local ML model using local training data and self-controlled compute (activity: trainMLModel; design option: train one complete ML model or train a part of the ML model). After local training is completed, *TraUpd* agents initiate the transmitInterimResult protocol (design option: gradients or parameter values) with their respective superordinate *ConCooSelUpd* agent as the respondent. The superordinate *ConCooSelUpd* agent waits until a defined threshold is reached (activity: awaitInterimResult; design option: waiting for a time-threshold or waiting for a response-threshold) and update their (part of the) ML model based on the interim results received (activity: updateMLModel; design option: batched update). The updates can be performed using algorithms such as FedAvg [63], FedProx [55], and FedBoost [34]. Each *ConCooSelUpd* agent can decide how often to repeat this training procedure with its descendant *TraUpd* agents. When the required number of training rounds is completed, *ConCooSelUpd* agents send the updated (part of the) ML model to their superordinate nodes (protocol: transmitInterimResult; design option: gradients or *parameter values*). Once the threshold of the root *ConCooSelUpd* agent is reached, a coalition-wide training round is completed. The procedure described for the operation phase is repeatedly executed until the dissolution phase is initiated by the root *ConCooSelUpd* agent.

*5.2.1 Key Traits.* The control archetype implements a strongly hierarchical organizational structure of agents and requires the coalition-wide synchronization of ML models. This leads to organizational structures in which a small fraction of agents control a CDML system. The control archetype relies on only one root *ConCooSelUpd* agent. If that agent crashes, the whole CDML system crashes [39, 81, 92]. Thus, the control archetype is commonly not crash-fault tolerant. The use of multiple *ConCooSelUpd* agents assigned to multiple layers of the hierarchy of the control archetype can make the system tolerant to crashes of single *ConCooSelUpd* agents [92, 104]. If one *ConCooSelUpd* agent crashes, other *ConCooSelUpd* agents can take the load of the crashed one. However, this redistribution of load to fewer *ConCooSelUpd* agents can drastically reduce the overall performance of the control archetype. The control archetype can be prone to performance bottlenecks due to a few central agents having to execute numerous computationally intensive activities and protocols [31, 92]. Such performance bottlenecks relate to computation [113] (i.e., during updating) and communication [113, 118] (i.e., sending and receiving interim results). The control archetype is prone to both, data poisoning attacks and inference attacks [45]. To mitigate these attacks, various defense measures were developed, such as secure aggregation [8]. Regarding the predictive performance of ML models trained collaboratively, the control archetype usually performs better than the confidentiality archetype [87]. The ML model usually converges faster than in CDML systems of the flexibility archetype [45]. The coalition can be dissolved after training because the coalition is not required to make ML model inferences.

*5.2.2 Variants of the Control Archetype.*

*TraUpd Agents as Tra Agents [31].* The *TraUpd* agents of the basic variant of the control archetype are exchanged with *Tra* agents. In this variant, interim results are only transmitted from *Tra* agents to *ConCooSelUpd* agents. No interim results are returned to *Tra* agents. *Tra* agents do not update their local ML models with interim results from other agents.

*ConCooSelUpd Agents as ConCooSelTraUpd Agents [17].* The *ConCooSelUpd* agents of the basic variant of the control archetype are exchanged with *ConCooSelTraUpd* agents. Agents on higher levels of the hierarchy have their own training data to train (parts of) the ML model themselves [17]. *ConCooSelTraUpd* agents train the ML model (activity: trainMLModel; design option: train one complete ML model or train a part of the ML model) while waiting for interim results of subordinate agents in the hierarchy.

*TraUpd Agents Train Two Complete ML Models [58]. TraUpd* agents can locally train two complete ML models (activity: `trainMLModel`; design option: `train two complete ML models`). *TraUpd* agents train one ML model on local data. The second ML model is trained on the first ML model. Only the gradients or parameter values resulting from the training of the second ML model are transmitted to the superordinate agent.

*5.2.3 Use Cases.* In industrial settings, CDML systems of the control archetype are utilized intra-organizational and inter-organizational. Intra-organizational settings include examples from the finance and manufacturing domains. In the finance domain, federated learning is utilized to develop ML models that support *company audits*. In this use case, the collaborating parties are subsidiaries of an audit company. In the manufacturing domain, federated learning is utilized to develop ML models that enable *predictive maintenance support for circulation pumps* and *quality inspection support for vehicle components*. In these use cases, the collaborating parties are fabrics of the same company. In such use cases, a central party controls the CDML system. The control archetype is often used in intra-organizational settings, probably, because its hierarchical structure resembles common organizational structures and, thus, simplifies the selection of a central party. The control archetype is used in inter-organizational settings related to healthcare and life sciences. In the healthcare domain, federated learning has been used to develop ML models that aid in *organ rejection detection*. The collaborating organizations are hospitals. In the life sciences domain, federated learning has been used to support clinical trials, with the collaborating organizations being hospitals and pharmaceutical companies. A challenge in these use cases is that organizations must agree on a central *ConCoolSelUpd* agent.

*5.2.4 Tools.* Several tools were developed to simulate, deploy, and combine CDML systems of the control archetype with PETs. For the *simulation* of such systems, *FedML* [37], *FEDn* [21], *Flower* [5], *NvideaFlare* [79], *OpenFL* [76], *PaddleFL* [70], and *Tensorflow Federated* [30] can be used. Tools that support *deployment* of CDML systems of the control archetype include *FedML* [37], *Tensorflow Federated* [30], *OpenFL* [76], *FATE* [20], *Flower* [5], *NvideaFlare* [79], *PaddleFL* [70], and *FEDn* [21]. Tools that help *combine CDML systems with PETs*, such as differential privacy, homomorphic encryption, and multi-party computation include *FATE* [20], *NvideaFlare* [79], *OpenFL* [76], *PaddleFL* [70], and *PySyft* [122].

## 5.3 Flexibility Archetype

The flexibility archetype is suitable for use cases with communication topologies that can change at run-time [113]. The flexibility archetype offers a high degree of agent autonomy. Agents can arbitrarily join and leave the flexibility archetype without impeding the functioning of the CDML system [113]. In its basic variant, each agent can select agents for collaboration. Moreover, agents can decide if and when they execute activities (e.g., `trainMLModel` or `updateMLModel`) and protocols (e.g., `signalReadiness` or `transmitInterimResult`). The flexibility archetype is weakly hierarchically organized. ML models are not synchronized coalition-wide during ML model training. Exemplary CDML systems of the flexibility archetype implement gossip learning [39], braintorrent [81], and decentralized federated learning [46, 52, 61, 106, 110, 113].

*Agent Model.* The flexibility archetype comprises the agent types *ConCooSelTraUpd* (i.e., the peer that submits the ML task to other peers) and *CooSelTraUpd* (i.e., all peers in the CDML system that do not submit ML tasks). In its basic configuration, the flexibility archetype comprises one *ConCooSelTraUpd* agent and at least one *CooSelTraUpd* agent.

*Acquaintance Model.* To train ML models, agents must establish a bidirectional communication path to at least one other agent (see Figure 3). Other agents can have the agent type *ConCooSelTraUpd* and *CooSelTraUpd*. Agents equitably decide with which agents they interact.

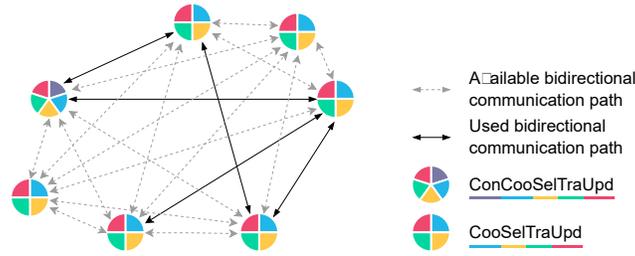

Fig. 3. Acquaintance model of the flexibility archetype for an exemplary training round

*Principal Functioning.* In the initialization phase, the *ConCooSelTraUpd* agent first defines the ML model (activity: defineInitialMLModel) and interim results (activity: defineInterimResult). Agents can join at any time (protocol: applyForCoalition; design option: always). The *ConCooSelTraUpd* agent distributes the ML model and the interim result definition to other agents in the CDML system (protocol: provideMLTask; design option: provide initial ML model definition and interim result definition).

In the operation phase, each *ConCooSelTraUpd* and *CooSelTraUpd* agent locally train the ML model. Agents that have completed the local training signal their readiness to activate their *updater* role for the upcoming training round (protocol: signalReadiness) and wait for other agents to signal their readiness (activity: awaitAgentReadiness). Then, at least one agent that signals its readiness is selected (activity: selectAgent) to receive interim results. Agents are usually selected randomly (design option: randomly, but can also be selected in a targeted manner (design option: based on agent attributes. The selection is announced to the selected agent (protocol: announceAgentSelection). Agents that are selected to activate the role *updater* wait (activity: awaitInterimResult) until they receive the interim results from other agents using the protocol transmitInterimResult (design option: gradients or parameter values). Lastly, the selected agents update their local ML model based on the interim results (activity: updateMLModel). The update can entail several interim results (design option: batched update) or only one interim result from another agent (design option: individual update). The updates can be performed using algorithms such as weighted averaging [38, 81], D-PSGD [57], and CHOCO-SGD [49].

This process is repeated until the dissolution phase is initiated. The flexibility archetype dissolves when no agents engage in collaborative training anymore.

*5.3.1 Key Traits.* The flexibility archetype is weakly hierarchical. ML models are not synchronized coalition-wide, and agents store different states of ML models. Agents have a high level of autonomy and, for example, can individually decide when to train collaboratively and with whom. Moreover, agents can individually decide to activate roles and execute activities and protocols, which leads to agents having little idle time [81]. This CDML archetype can offer the highest degree of decentralization compared to the other CDML archetypes.

The flexibility archetype can handle agent crashes better than the control archetype [39]. An agent dropping out of the system may temporarily reduce the performance of the flexibility archetype, but the flexibility archetype can recover because a new agent can be easily integrated into the training process [45]. Because agents can operate largely autonomously and are replaceable, no central agent is focal for the proper functioning of CDML systems of the flexibility archetype. However, such replacement may not always be possible because the flexibility archetype does not require redundant agents.

Because there is no coalition-wide synchronization of ML models, improvement of ML models can be slow and ML models may not be of high accuracy. "Catastrophic forgetting" [81], can occur when each ML model update replaces previously acquired ML model parameters with low generalization error, with new ML model parameters with higher generalization error [81].

The flexibility archetype is commonly not robust against malicious agents. Malicious agents can conduct data poisoning attacks [73]. Malicious agents can obfuscate their identities by arbitrarily joining and dropping out of the CDML system and arbitrarily switching their collaboration partners. Such obfuscation can facilitate agents in performing malicious activities without being detected [e.g., because selecting non-malicious agents for a training round might be impeded by ineffective reputation systems; 44]. Moreover, even when malicious agents are identified, it is hard to punish them because rules (e.g., agents that act maliciously are forced to leave the system) are hardly enforceable in the flexibility archetype. The coalition can be dissolved after ML model training because the CDML system is not required to make ML model inferences.

#### 5.3.2 Variants of the Flexibility Archetype.

*Additional CooSel Agent [39].* There can be a dedicated *CooSel* agent [39]. The remaining agents do not have the *selector* role and become *ConCooTraUpd* and *ConTraUpd* agents. In each training round, the *CooSel* agent selects a subset of the *ConCooTraUpd* and *CooTraUpd* agents to activate the updater role (activity: selectAgent; design option: randomly) and assigns each of the remaining agents to one of the agents with an activated *updater* role. Each agent then sends its interim result to the agent it was assigned to (protocol: transmitInterimResult; design option: gradients or parameter values).

#### 5.3.3 Use Cases.
We could not identify industrial uses of the flexibility archetype. Researchers proposed gossip learning for intra-organizational settings, such as large-scale device networks, to help filter spam [68]. Braintorrent was proposed for conducting brain segmentations for inter-organizational settings involving medical centers [81].

#### 5.3.4 Tools.
GoLF [67] and GossipFL [86] can help *simulate* CDML systems of the flexibility archetype.

### 5.4 Robustness Archetype

The robustness archetype is suitable for use cases where agents can drop out of the coalition during ML model training (e.g., due to crashes or network failures). A large fraction of agents is redundant and, thus, can replace each other. The robustness archetype is weakly hierarchically organized and performs coalition-wide synchronization of ML models. Exemplary CDML systems of the robustness archetype are blockchain-based CDML systems [47, 53, 56, 82, 97].

*Agent Model.* The robustness archetype comprises the agent types *ConCooSelTraUpd* (e.g., the agent that submits an ML task in swarm learning systems [97]) and *CooSelTraUpd* (e.g., agents that do not submit an ML task in swarm learning systems). In its basic configuration, the robustness archetype must comprise one *ConCooSelTraUpd* agent and at least one *CooSelTraUpd* agent, to redundantly act in the roles *coordinator*, *selector*, *trainer*, and *updater*.

*Acquaintance Model.* CDML systems of the robustness archetype can have bidirectional communication paths between all *ConCooSelTraUpd* agents and *CooSelTraUpd* agents (see Figure 4).

*Principal Functioning.* In the initialization phase, the *ConCooSelTraUpd* agent defines an ML model and interim results and distributes the definitions to other agents (protocol: provideMLTask; design option: provide ML model definition and interim result definition). *CooSelTraUpd* agents can join at any time (protocol: applyForCoalition; design option: always).

In the operation phase, *ConCooSelTraUpd* and *CooSelTraUpd* agents broadcast their readiness to activate their roles *updater* and *trainer* (protocol: signalReadiness). All agents that receive the broadcast individually decide whether other agents should activate the *trainer* and *updater* role. Agents broadcast their decisions to all agents in the CDML system. The final selection of agents that activate the roles *trainer* and *updater* is made through a consensus mechanism (activity: selectAgent; design option: based on votes from other agents). For example, all agents vote on which agents should update the global ML model in the subsequent training round. The agent with the most votes is selected to activate the *updater* role. Next, *ConCooSelTraUpd* and *CooSelTraUpd* agents start training the ML model (activity: trainMLModel; design option: train a complete ML Model). All agents selected to activate the *updater* role receive identical interim results from agents that trained their ML model (protocol: transmitInterimResult; design option: gradients or parameter values). All agents use identical interim results to update the ML model (activity: updateMLModel). For the update, all selected *updater* agents use the results of from all other agents (design option: batched update). Updates can be performed using algorithms like weighted averaging [82], distributed approximate newton [112], and softmax aggregation [103]. All agents that computed ML model updates broadcast their new interim results to all agents in the CDML system (protocol: transmitInterimResult).

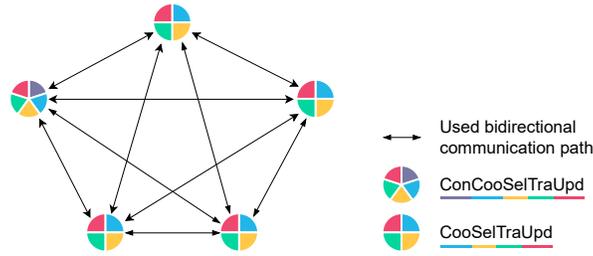

Fig. 4. Exemplary acquaintance model of the robustness archetype

This process is repeated until the start of the dissolution phase. The dissolution phase starts when no agents engage in the collaborative training anymore.

*5.4.1 Key Traits.* The robustness archetype is weakly hierarchical and uses coalition-wide synchronization to train global ML models. Identical agent types are assigned to multiple agents. Agents process and store data of the global ML model redundantly, increasing robustness of this archetype.

The robustness archetype uses a fully connected communication network [113] and does not rely on a central agent. This design protects the robustness archetype from issues caused by agent drop-outs [56].

The robustness archetype can be prone to data poisoning attacks [11]. Due to the absence of a central agent that coordinates the CDML system, implementing common defenses against poisoning attacks can be challenging. Approaches to defend against poisoning attacks include incentive mechanisms and reputation systems [119]. Adding client-level differential privacy can protect the robustness archetype against inference attacks to achieve similar training data confidentiality like federated learning systems [45].

The robustness archetype allows for the replacement of *updater* agents after each training round. Agents in the robustness archetype usually require large computational resources, for example, to compute ML model updates based on interim results from all other agents in the CDML system [113]. The coalition is not required to make ML model inferences and can be dissolved after training.

*5.4.2 Variants of the Robustness Archetype.*

*A subset of agents activates the updater role per training round [56, 82, 97].* Interim results are transmitted to and stored by all agents, but only a subset of agents activates the *updater* role. From all *ConCooSelTraUpd* and *CooSelTraUpd* agents that signal their readiness (protocol: *signalReadiness*), not all agents are selected (activity: selectAgent; design options: based on agent attributes, based on votes from other agents, or randomly) to activate their *updater* role in every training round. In some cases, only one agent is selected to activate its *updater* role [97].

*5.4.3 Use Cases.* In industrial settings, CDML systems of the robustness archetype are predominantly utilized in inter-organizational settings. For example, in the healthcare domain, hospitals used swarm learning to develop ML models that aid in *disease diagnostic*. The hospitals aimed to collaboratively develop an ML model, with each organization having equal rights and responsibilities—CDML system characteristics promoted by the robustness archetype.

*5.4.4 Tools.* For the *simulation* of CDML systems of the robustness archetype, *DFL* [88] can be used. Tools that support *deployment* of CDML systems of the robustness archetype include *swarm learning library* [40] and *DFL* [88].

## 6 Discussion

This section describes principal findings and outlines contributions to research and practice. The limitations of this work are then discussed, followed by three proposed avenues for future research.

### 6.1 Principal Findings

This work presents a CDML system conceptualization that describes the principal functioning of CDML systems involving five agent roles (i.e., *configurator*, *coordinator*, *selector*, *trainer*, and *updater*), and respective activities (e.g., updateMLModel) and protocols (e.g., transmitInterimResult). Moreover, we present 33 design options to customize CDML systems, including agent types (e.g., *TraUpd*) and different communication paths. Based on common combinations of design options, we developed four CDML archetypes (i.e., the confidentiality archetype, control archetype, flexibility archetype, and robustness archetype) with different key traits.

We recognized a substantial increase in the variety of CDML system designs through contributions from practice and research. After federated learning systems, CDML systems, such as split learning systems and swarm learning systems, have moved into the focus of practice and research. Moreover, various tools were developed to simulate and deploy such CDML systems. We observed that for CDML systems of the control archetype, particularly federated learning systems, there are numerous industry uses and a wide range of tools available to developers. In contrast, there are comparatively few applications and tools for other CDML archetypes.

We observed strong similarities among CDML system designs. Assisted learning systems and split learning systems use similar design options; for example, they comprise only *ConCooSelUpd* and *ConTraUpd* agents. Moreover, swarm learning systems and blockchain-based decentralized federated learning systems use similar design options. For example, both implement the agent types *ConCooSelTraUpd* and *CooSelTraUpd* but differ regarding the number of agents with an active *updater* role in each training round.

The presented CDML archetypes and their key traits show that no one-size-fits-all CDML system can be used for every use case. Developers must carefully assess the suitability of CDML systems for use cases based on different traits. For instance, the redundant distribution of roles in swarm learning enhances robustness. However, in use cases where most agents have few resources, mandating that all agents perform all roles may exceed agents' resource capacities. Redundancy in agent roles can be better suited for use cases characterized by likely agent drop-outs.

In the agent model (see Section 4.2.3), we present the agent types that we identified in the analyzed publications. The presented agent types represent a subset of the possible combinations of agent roles. For example, we did not identify agents that comprise only the roles *trainer* and *selector* in the analyzed CDML systems, even though the implementation of such agents could be possible as long as all roles are distributed to agents in CDML systems. CDML systems that assign each agent only one role could also have new traits, including agents requiring fewer resources, that might be useful in many use cases. Because of the theoretical possibility of more agent types and combinations of design options, more CDML system designs with different traits may become available in the future.

## 6.2 Contributions to Practice and Research

With this study, we contribute to practice and research in three principal ways. First, by presenting the CDML conceptualization, we support a better understanding of the principal functioning and design differences of CDML systems. This enables systematic comparisons between CDML system designs covering a large set of design options, which helps develop CDML systems that meet use case requirements. The roles model and the interaction model of the CDML conceptualization present the main design commonalities of CDML systems (e.g., the use of specific agent roles and the principal training process). The three agent-based models on the design level (i.e., agent model, service model, and acquaintance model) can guide the systematic comparison between CDML system designs and the customization of CDML system designs to meet use case requirements. Moreover, the developed agent-based models can facilitate the application of the Gaia methodology for developing custom CDML system designs.

Second, by showcasing CDML archetypes, we offer starting points for combining design options commonly used in practice and research to develop CDML system designs with different key traits. The CDML archetypes can be customized using the design options presented in the CDML conceptualization to develop blueprints of CDML systems. Thereby, in combination, the CDML archetypes and the CDML conceptualization offer actionable help in guiding the design of CDML systems that suit use cases.

Third, by presenting key traits of CDML archetypes, we support developers in deciding on combinations of design options to meet use case requirements. The key traits of CDML archetypes enable developers to choose the most fitting CDML archetype for use cases. Using the selected CDML archetype as a starting point, developers can use the CDML conceptualization and customize the archetype to show additional required traits. By executing this process, developers can evaluate CDML system designs in terms of their suitability for use cases before implementation.

## 6.3 Limitations

For the development of the CDML system conceptualization, the CDML archetypes, and the identification of key traits, we analyzed publications and CDML systems representative of the CDML field. With our selection of publications and CDML systems for analysis, we aimed to cover the large spectrum of different CDML system designs. However, the number of publications and CDML systems significantly increased in the past years, making it impossible to incorporate all publications in our study with methodological rigor but only a representative set of publications. Thus, the CDML conceptualization may not cover all CDML system designs.

We strove to extract and understand key design aspects of CDML systems (e.g., activities, processes, and roles), requiring the resolution of ambiguities, and to set extracted insights in relationships (e.g., roles and responsibilities). Although well-suited to conduct such research, qualitative research is inherently prone to subjective biases, for example, because publications are individually interpreted depending on personal conceptions. Despite our efforts to reduce such biases (e.g., through feedback on our results from ML experts and orientation toward the Gaia methodology), we cannot guarantee that we have eliminated them. Other researchers may bring forth different CDML system conceptualizations.

The analyzed publications mainly focus on core training processes in CDML systems [31, 39, 81, 90, 113]. Other system components required to operate CDML systems are mostly neglected. By triangulating descriptions of CDML systems based on our coding and intense discussions with ML experts, we aimed to complete fragmented descriptions of CDML systems. Still, the CDML conceptualization may lack aspects not specifically mentioned in the analyzed publications. Similarly, some of the examined publications lack sufficient detail in their descriptions of permissions of roles, activities, and protocols. This hindered us in describing permissions associated with agent roles and impeded the development of a complete service model as demanded by the Gaia methodology. Instead, we developed a preliminary service model to offer a starting point for the design and implementation of activities and protocols.

### 6.4 Future Research

The CDML conceptualization offers a foundation for knowledge transfers within the CDML community (e.g., to develop new CDML systems) and from other disciplines (e.g., from game theory to develop incentive mechanism [89]). In the following, we describe four areas for knowledge transfer that, from our perspective, are particularly interesting for improving CDML systems in future research.

*Hyperparameter Optimization.* Automated hyperparameter optimization (HPO) has become very important in the development of ML models for manifold purposes [36], such as to improve ML model performance and decrease necessary computations in the training of ML models. For most automated HPO methods, such as Bayesian optimization [9, 27, 83], the availability of complete training data sets is assumed. This assumption lies at odds with decentralized training data management in CDML systems. Extant automated HPO methods are hardly applicable to CDML systems, which may result in under-optimized ML models trained in CDML systems [36]. The CDML conceptualization can be a foundation for future research to identify challenges in performing HPO in CDML systems with different designs and develop corresponding solutions.

*Data Confidentiality.* The exchange of interim results instead of training data does not guarantee training data confidentiality per se [7]. To protect training data confidentiality, combining CDML and PETs, such as differential privacy and homomorphic encryption, has become promising [98, 115]. Future research should develop guidelines for reasonably combining the CDML paradigm with other PETs.

*Robustness.* Agents may pursue individual goals in CDML systems. However, ensuring the accurate alignment between individual agent goals and the overarching goal of the CDML system is critical. Misalignment can have detrimental consequences, such as the introduction of the free-rider problem [25] and incentivizing agents to poison training data or ML models [4, 89, 114]. The free-rider problem is characterized by agents that provide subpar data while being able to improve their ML model from interim results received from other agents. Integrating robustness measures from diverse fields into CDML systems, such as financial incentives in economics and normative principles in sociology for agent behavior coordination [42, 44, 48, 89], could enhance the robustness of CDML systems against challenges, such as anticipating malicious actions of agents in CDML systems. Future research should extend the CDML conceptualization to include design options that improve the robustness of CDML systems and protect ML model training from malicious agent activity.

## 7 Conclusion

Based on agent-oriented modeling, this work presents a CDML conceptualization that supports a better understanding of CDML systems and can inform comparison of CDML system designs in terms of their suitability for use cases. Based on the CDML conceptualization, we developed four CDML archetypes with different key traits to guide developers in designing CDML systems with desired characteristics. Given the variety of CDML system designs presented in this work, we believe that CDML has reached a level of maturity that can help increase the data sovereignty of private and commercial data subjects in the context of providing ML training data. This could usher in a new era of ML development that relies less on single hyperscalers and more on collective and decentralized work. We hope that the CDML conceptualization will support development of CDML systems suitable for use cases [e.g., by facilitating the use of the Gaia methodology; 102] to enable the training of ML models on sufficient (even sensitive) training data at a large scale. Owing to the considerable attention that CDML systems have garnered in practice and research and the continuing development of novel CDML concepts beyond federated learning, we encourage the advancement of the CDML conceptualization in the future.

## Acknowledgments

We thank Anne Hüsges, Benjamin Sturm, Kathrin Brecker, Marc Zöller, Mikael Beyene, Richard Guse, Simon Warsinsky, and Tobias Dehling for their valuable feedback on this work. This work was supported by funding from the topic Engineering Secure Systems of the Helmholtz Association (HGF) and by KASTEL Security Research Labs.

## References

[1] George J. Annas. 2003. HIPAA Regulations–A New Era of Medical-Record Privacy? *New England Journal of Medicine* 348, 15 (2003), 1486–1490. https://doi.org/10.1056/NEJMlim035027

[2] Rezak Aziz, Soumya Banerjee, Samia Bouzefrane, and Thinh Le Vinh. 2023. Exploring Homomorphic Encryption and Differential Privacy Techniques towards Secure Federated Learning Paradigm. *Future Internet* 15, 9 (2023). https://doi.org/10.3390/fi15090310

[3] Aayushi Bansal, Rewa Sharma, and Mamta Kathuria. 2022. A Systematic Review on Data Scarcity Problem in Deep Learning: Solution and Applications. *Comput. Surveys* 54, 10 (2022). https://doi.org/10.1145/3502287

[4] Xianglin Bao, Cheng Su, Yan Xiong, Wenchao Huang, and Yifei Hu. 2019. FLChain: A Blockchain for Auditable Federated Learning with Trust and Incentive. In *2019 5th International Conference on Big Data Computing and Communications*. QingDao, China, 151–159. https://doi.org/10.1109/BIGCOM.2019.00030

[5] Daniel J. Beutel, Taner Topal, Akhil Mathur, Xinchi Qiu, et al. 2022. Flower: A Friendly Federated Learning Research Framework. https://doi.org/10.48550/arXiv.2007.14390

[6] Michael Blot, David Picard, Matthieu Cord, and Nicolas Thome. 2016. Gossip training for deep learning. http://arxiv.org/abs/1611.09726

[7] Franziska Boenisch, Adam Dziedzic, Roei Schuster, Ali Shahin Shamsabadi, et al. 2023. When the curious abandon honesty: Federated learning is not private. In *2023 IEEE 8th European Symposium on Security and Privacy (EuroS&P)*. IEEE, 175–199. https://doi.org/10.1109/EuroSP57164.2023.00020

[8] Keith Bonawitz, Hubert Eichner, Wolfgang Grieskamp, Dzmitry Huba, et al. 2019. Towards Federated Learning at Scale: System Design. https://arxiv.org/abs/1902.01046

[9] Eric Brochu, Vlad M. Cora, and Nando de Freitas. 2010. A Tutorial on Bayesian Optimization of Expensive Cost Functions, with Application to Active User Modeling and Hierarchical Reinforcement Learning. https://doi.org/10.48550/arXiv.1012.2599

[10] Cheng Chen, Jiaying Zhou, Jie Ding, and Yi Zhou. 2022. Assisted Learning for Organizations with Limited Data. https://arxiv.org/abs/2109.09307

[11] Kongyang Chen, Huaiyuan Zhang, Xiangyu Feng, Xiaoting Zhang, Bing Mi, and Zhiping Jin. 2023. Backdoor attacks against distributed swarm learning. *ISA Transactions* 141 (2023), 59–72. https://doi.org/10.1016/j.isatra.2023.03.034

[12] Xing Chen, Jingtao Li, and Chaitali Chakrabarti. 2021. Communication and Computation Reduction for Split Learning using Asynchronous Training. In *2021 IEEE Workshop on Signal Processing Systems*. Coimbra, Portugal, 76–81. https://doi.org/10.1109/SiPS52927.2021.00022

[13] Juliet Corbin and Anselm Strauss. 2015. *Basics of qualitative research: techniques and procedures for developing grounded theory* (4 ed.). SAGE Publications, Los Angeles, CA, USA.

[14] Enmao Diao. 2020. *GAL: Gradient Assisted Learning for Decentralized Multi-Organization Collaborations*. Retrieved 2024-08-05 from https://github.com/diaoenmao/GAL-Gradient-Assisted-Learning-for-Decentralized-Multi-Organization-Collaborations

[15] Enmao Diao, Jie Ding, and Vahid Tarokh. 2021. HeteroFL: Computation and Communication Efficient Federated Learning for Heterogeneous Clients. https://arxiv.org/abs/2010.01264

[16] Enmao Diao, Jie Ding, and Vahid Tarokh. 2022. GAL: Gradient Assisted Learning for Decentralized Multi-Organization Collaborations. http://arxiv.org/abs/2106.01425

[17] Enmao Diao, Jie Ding, and Vahid Tarokh. 2022. SemiFL: Semi-Supervised Federated Learning for Unlabeled Clients with Alternate Training. In *Advances in Neural Information Processing Systems*, Vol. 35. Curran Associates, Inc., New Orleans, LA, USA, 17871–17884.

[18] Enmao Diao, Jie Ding, and Vahid Tarokh. 2023. *Collaborative Data Analysis for All*. Retrieved 2024-08-09 from https://github.com/Collaborative-AI/colda

[19] Michael Diskin, Alexey Bukhtiyarov, Max Ryabinin, Lucile Saulnier, et al. 2021. Distributed deep learning in open collaborations. *Advances in Neural Information Processing Systems* 34 (2021), 7879–7897.

[20] Federated AI Ecosystem. 2019. *FATE: An Industrial Grade Federated Learning Framework*. Retrieved 2024-08-05 from https://github.com/FederatedAI/FATE
[21] Morgan Ekmefjord, Addi Ait-Mlouk, Sadi Alawadi, Mattias Akesson, et al. 2022. Scalable federated machine learning with FEDn. In *2022 22nd IEEE International Symposium on Cluster, Cloud and Internet Computing (CCGrid)*. IEEE, Taormina, Italy, 555–564. https://doi.org/10.1109/CCGrid54584.2022.00065
[22] Ege Erdoğan, Alptekin Küpçü, and A. Ercüment Çiçek. 2022. UnSplit: Data-Oblivious Model Inversion, Model Stealing, and Label Inference Attacks against Split Learning. In *Proceedings of the 21st Workshop on Privacy in the Electronic Society*. New York, NY, USA, 115–124. https://doi.org/10.1145/3559613.3563201
[23] European Parliament and Council of the European Union. 2016. Regulation (EU) 2016/679 of the European Parliament and of the Council. https://data.europa.eu/eli/reg/2016/679/oj
[24] Centers for Medicare & Medicaid Services. 1996. The Health Insurance Portability and Accountability Act of 1996 (HIPAA). http://www.cms.hhs.gov/hipaa/
[25] Yann Fraboni, Richard Vidal, and Marco Lorenzi. 2020. Free-rider Attacks on Model Aggregation in Federated Learning. https://arxiv.org/abs/2006.11901
[26] Saurabh Gajbhiye, Priyanka Singh, and Shaifu Gupta. 2022. Data poisoning attack by label flipping on splitfed learning. In *International Conference on Recent Trends in Image Processing and Pattern Recognition*. 391–405.
[27] Roman Garnett. 2023. *Bayesian optimization*. Cambridge University Press, Cambridge, United Kingdom.
[28] Jonas Geiping, Hartmut Bauermeister, Hannah Dröge, and Michael Moeller. 2020. Inverting gradients-how easy is it to break privacy in federated learning? *Advances in neural information processing systems* 33 (2020), 16937–16947.
[29] Maoguo Gong, Yuanqiao Zhang, Yuan Gao, A. K. Qin, et al. 2024. A Multi-Modal Vertical Federated Learning Framework Based on Homomorphic Encryption. *IEEE Transactions on Information Forensics and Security* 19 (2024), 1826–1839. https://doi.org/10.1109/TIFS.2023.3340994
[30] Google. 2018. *TensorFlow Federated*. Retrieved 2024-07-15 from https://github.com/google-parfait/tensorflow-federated
[31] Neel Guha, Ameet Talwalkar, and Virginia Smith. 2019. One-Shot Federated Learning. https://arxiv.org/abs/1902.11175
[32] Otkrist Gupta and Ramesh Raskar. 2018. Distributed learning of deep neural network over multiple agents. *Journal of Network and Computer Applications* 116 (2018), 1–8. https://doi.org/10.1016/j.jnca.2018.05.003
[33] Alon Halevy, Peter Norvig, and Fernando Pereira. 2009. The Unreasonable Effectiveness of Data. *IEEE Intelligent Systems* 24, 2 (2009), 8–12. https://doi.org/10.1109/MIS.2009.36
[34] Jenny Hamer, Mehryar Mohri, and Ananda Theertha Suresh. 2020. FedBoost: A Communication-Efficient Algorithm for Federated Learning. In *Proceedings of the 37th International Conference on Machine Learning (Proceedings of Machine Learning Research, Vol. 119)*. PMLR, 3973–3983.
[35] Andrew Hard, Chloé M Kiddon, Daniel Ramage, Francoise Beaufays, et al. 2018. Federated Learning for Mobile Keyboard Prediction. https://arxiv.org/abs/1811.03604
[36] Niklas Hasebrook, Felix Morsbach, Niclas Kannengießer, Marc Zöller, et al. 2023. Practitioner Motives to Select Hyperparameter Optimization Methods. https://arxiv.org/abs/2203.01717v2
[37] Chaoyang He, Songze Li, Jinhyun So, Xiao Zeng, et al. 2020. FedML: A Research Library and Benchmark for Federated Machine Learning. https://doi.org/arXiv:2007.13518
[38] István Hegedüs, Gábor Danner, and Márk Jelasity. 2021. Decentralized learning works: An empirical comparison of gossip learning and federated learning. *J. Parallel and Distrib. Comput.* 148 (2021), 109–124. https://doi.org/10.1016/j.jpdc.2020.10.006
[39] István Hegedűs, Gábor Danner, and Márk Jelasity. 2019. Gossip Learning as a Decentralized Alternative to Federated Learning. In *Distributed Applications and Interoperable Systems*, José Pereira and Laura Ricci (Eds.). Vol. 11534. Springer International Publishing, Cham, Switzerland, 74–90. https://doi.org/10.1007/978-3-030-22496-7_5
[40] HewlettPackard. 2022. *A simplified library for decentralized, privacy preserving machine learning*. Retrieved 2024-08-09 from https://github.com/HewlettPackard/swarm-learning
[41] Yaochen Hu, Di Niu, Jianming Yang, and Shengping Zhou. 2019. FDML: A Collaborative Machine Learning Framework for Distributed Features. In *Proceedings of the 25th ACM SIGKDD International Conference on Knowledge Discovery & Data Mining*. Anchorage, AK, USA, 2232–2240. https://doi.org/10.1145/3292500.3330765
[42] Laura Hunter and Erin Leahey. 2008. Collaborative Research in Sociology: Trends and Contributing Factors. *The American Sociologist* 39, 4 (2008), 290–306. https://doi.org/10.1007/s12108-008-9042-1
[43] Weiwei Jiang, Haoyu Han, Yang Zhang, and Jianbin Mu. 2024. Federated split learning for sequential data in satellite–terrestrial integrated networks. *Information Fusion* 103 (2024), 102141. https://doi.org/10.1016/j.inffus.2023.102141
[44] David Jin, Niclas Kannengießer, Benjamin Sturm, and Ali Sunyaev. 2022. Tackling Challenges of Robustness Measures for Agent Collaboration in Open Multi-Agent Systems. In *55th Hawaii International Conference on System Sciences*. Honolulu, HI, USA, 7585–7594. https://doi.org/10.24251/HICSS.2022.911

[45] Peter Kairouz, H. Brendan McMahan, Brendan Avent, Aurélien Bellet, et al. 2021. Advances and Open Problems in Federated Learning. https://doi.org/10.48550/arXiv.1912.04977

[46] Shivam Kalra, Junfeng Wen, Jesse C. Cresswell, Maksims Volkovs, and H. R. Tizhoosh. 2023. Decentralized federated learning through proxy model sharing. *Nature Communications* 14, 1 (2023), 2899. https://doi.org/10.1038/s41467-023-38569-4

[47] Hyesung Kim, Jihong Park, Mehdi Bennis, and Seong-Lyun Kim. 2020. Blockchained On-Device Federated Learning. *IEEE Communications Letters* 24, 6 (2020), 1279–1283. https://doi.org/10.1109/LCOMM.2019.2921755

[48] Daniel Kirste, Niclas Kannengießer, Ricky Lamberty, and Ali Sunyaev. 2023. How Automated Market Makers Approach the Thin Market Problem in Cryptoeconomic Systems. https://doi.org/10.48550/arXiv.2309.12818

[49] Anastasia Koloskova, Sebastian Stich, and Martin Jaggi. 2019. Decentralized Stochastic Optimization and Gossip Algorithms with Compressed Communication. In *Proceedings of the 36th International Conference on Machine Learning (Proceedings of Machine Learning Research, Vol. 97)*. PMLR. https://proceedings.mlr.press/v97/koloskova19a.html

[50] Jakub Konečný, Brendan McMahan, and Daniel Ramage. 2015. Federated Optimization:Distributed Optimization Beyond the Datacenter. http://arxiv.org/abs/1511.03575

[51] Tim Kraska, Ameet Talwalkar, John C Duchi, Rean Griffith, Michael J Franklin, and Michael I Jordan. 2013. MLbase: A Distributed Machine-learning System. In *Cidr*, Vol. 1. 2–1.

[52] Chengxi Li, Gang Li, and Pramod K. Varshney. 2022. Decentralized Federated Learning via Mutual Knowledge Transfer. *IEEE Internet of Things Journal* 9, 2 (2022), 1136–1147. https://doi.org/10.1109/JIOT.2021.3078543

[53] Jun Li, Yumeng Shao, Kang Wei, Ming Ding, et al. 2021. Blockchain Assisted Decentralized Federated Learning (BLADE-FL): Performance Analysis and Resource Allocation. https://doi.org/10.48550/arxiv.2101.06905

[54] Mu Li, David G Andersen, Jun Woo Park, Alexander J Smola, et al. 2014. Scaling distributed machine learning with the parameter server. In *11th USENIX Symposium on Operating Systems Design and Implementation*. Broomfield, CO, USA, 583–598.

[55] Tian Li, Anit Kumar Sahu, Manzil Zaheer, Maziar Sanjabi, Ameet Talwalkar, and Virginia Smith. 2020. Federated Optimization in Heterogeneous Networks. http://arxiv.org/abs/1812.06127

[56] Yuzheng Li, Chuan Chen, Nan Liu, Huawei Huang, et al. 2021. A Blockchain-Based Decentralized Federated Learning Framework with Committee Consensus. *IEEE Network* 35, 1 (2021), 234–241. https://doi.org/10.1109/MNET.011.2000263

[57] Xiangru Lian, Ce Zhang, Huan Zhang, Cho-Jui Hsieh, et al. 2017. Can decentralized algorithms outperform centralized algorithms? a case study for decentralized parallel stochastic gradient descent. In *Proceedings of the 31st International Conference on Neural Information Processing Systems (NIPS'17)*. Curran Associates Inc., Red Hook, NY, USA, 5336–5346. Long Beach, California, USA.

[58] Paul Pu Liang, Terrance Liu, Liu Ziyin, Nicholas B. Allen, et al. 2020. Think Locally, Act Globally: Federated Learning with Local and Global Representations. https://arxiv.org/abs/2001.01523

[59] Zheng Lin, Guangyu Zhu, Yiqin Deng, Xianhao Chen, Yue Gao, Kaibin Huang, and Yuguang Fang. 2024. Efficient Parallel Split Learning over Resource-constrained Wireless Edge Networks. *IEEE Transactions on Mobile Computing* (2024), 1–16. https://doi.org/10.1109/TMC.2024.3359040

[60] Ji Liu, Jizhou Huang, Yang Zhou, Xuhong Li, Shilei Ji, Haoyi Xiong, and Dejing Dou. 2022. From distributed machine learning to federated learning: a survey. *Knowledge and Information Systems* 64, 4 (2022), 885–917.

[61] Wei Liu, Li Chen, and Wenyi Zhang. 2022. Decentralized Federated Learning: Balancing Communication and Computing Costs. *IEEE Transactions on Signal and Information Processing over Networks* 8 (2022), 131–143. https://doi.org/10.1109/TSIPN.2022.3151242

[62] Guodong Long, Yue Tan, Jing Jiang, and Chengqi Zhang. 2020. Federated learning for open banking. In *Federated Learning: Privacy and Incentive*. Springer, Heidelberg, Germany, 240–254.

[63] Brendan McMahan, Eider Moore, Daniel Ramage, Seth Hampson, and Blaise Aguera y Arcas. 2017. Communication-Efficient Learning of Deep Networks from Decentralized Data. In *Proceedings of the 20th International Conference on Artificial Intelligence and Statistics*, Aarti Singh and Jerry Zhu (Eds.), Vol. 54. 1273–1282.

[64] Payman. Mohassel and Peter Rindal. 2018. ABY3: A mixed protocol framework for machine learning. In *Proceedings of the 2018 ACM SIGSAC Conference on Computer and Communications Security*. New York, NY, USA, 35–52. https://doi.org/10.1145/3243734.3243760

[65] Payman Mohassel and Yupeng Zhang. 2017. Secureml: A system for scalable privacy-preserving machine learning. In *2017 IEEE symposium on security and privacy (SP)*. IEEE, 19–38.

[66] Olga Ohrimenko, Felix Schuster, Cédric Fournet, Aastha Mehta, Sebastian Nowozin, Kapil Vaswani, and Manuel Costa. 2016. Oblivious {Multi-Party} machine learning on trusted processors. In *25th USENIX Security Symposium (USENIX Security 16)*. 619–636.

[67] Robert Ormandi. 2011. *Gossip Learning Framework (GoLF)*. Retrieved 2024-07-28 from https://github.com/ormandi/Gossip-Learning-Framework

[68] Róbert Ormándi, István Hegedüs, and Márk Jelasity. 2013. Gossip Learning with Linear Models on Fully Distributed Data. *Concurrency and Computation: Practice and Experience* 25, 4 (2013), 556–571. https://doi.org/10.1002/cpe.2858

[69] Ahmet Murat Ozbayoglu, Mehmet Ugur Gudelek, and Omer Berat Sezer. 2020. Deep learning for financial applications: A survey. *Applied Soft Computing* 93 (2020), 106384. https://doi.org/10.1016/j.asoc.2020.106384

[70] PaddlePaddle. 2020. *PaddleFL: Federated Deep Learning in PaddlePaddle*. Retrieved 2024-08-11 from https://github.com/PaddlePaddle/PaddleFL

[71] Dario Pasquini, Giuseppe Ateniese, and Massimo Bernaschi. 2021. Unleashing the Tiger: Inference Attacks on Split Learning. In *Proceedings of the 2021 ACM SIGSAC Conference on Computer and Communications Security*. Association for Computing Machinery, New York, NY, USA, 2113–2129. https://doi.org/10.1145/3460120.3485259

[72] Diego Peteiro-Barral and Bertha Guijarro-Berdiñas. 2013. A survey of methods for distributed machine learning. *Progress in Artificial Intelligence* 2, 1 (2013), 1–11.

[73] Alexandre Pham, Maria Potop-Butucaru, Sébastien Tixeuil, and Serge Fdida. 2024. Data Poisoning Attacks in Gossip Learning. http://arxiv.org/abs/2403.06583

[74] Foster J. Provost and Daniel N. Hennessy. 1996. Scaling up: Distributed machine learning with cooperation. *Proceedings of the Thirteenth National Conference on Artificial Intelligence and Eighth Innovative Applications of Artificial Intelligence Conference (AAAI/IAAI'96)* Vol. 1. (1996), 74–79.

[75] Samyam Rajbhandari, Jeff Rasley, Olatunji Ruwase, and Yuxiong He. 2020. Zero: Memory optimizations toward training trillion parameter models. In *SC20: International Conference for High Performance Computing, Networking, Storage and Analysis*. Atlanta, GA, USA, 1–16.

[76] G. Anthony Reina, Alexey Gruzdev, Patrick Foley, Olga Perepelkina, et al. 2022. OpenFL: An open-source framework for Federated Learning. https://doi.org/10.1088/1361-6560/ac97d9

[77] Hanchi Ren, Jingjing Deng, and Xianghua Xie. 2022. Grnn: generative regression neural network—a data leakage attack for federated learning. *ACM Transactions on Intelligent Systems and Technology (TIST)* 13, 4 (2022), 1–24.

[78] Nicola Rieke, Jonny Hancox, Wenqi Li, Fausto Milletarì, et al. 2020. The future of digital health with federated learning. *npj Digital Medicine* 3, 119 (2020). https://doi.org/10.1038/s41746-020-00323-1

[79] Holger R. Roth, Yan Cheng, Yuhong Wen, Isaac Yang, et al. 2022. NVIDIA FLARE: Federated Learning from Simulation to Real-World. https://doi.org/10.48550/ARXIV.2210.13291

[80] Jennifer Rowley. 2012. Conducting research interviews. *Management Research Review* 35, 3/4 (March 2012), 260–271.

[81] Abhijit Guha Roy, Shayan Siddiqui, Sebastian Pölsterl, Nassir Navab, and Christian Wachinger. 2019. BrainTorrent: A Peer-to-Peer Environment for Decentralized Federated Learning. https://arxiv.org/abs/1905.06731

[82] Oliver Lester Saldanha, Philip Quirke, Nicholas P. West, Jacqueline A. James, et al. 2022. Swarm learning for decentralized artificial intelligence in cancer histopathology. *Nature Medicine* 28, 6 (June 2022), 1232–1239. https://doi.org/10.1038/s41591-022-01768-5

[83] Bobak Shahriari, Kevin Swersky, Ziyu Wang, Ryan P. Adams, and Nando De Freitas. 2016. Taking the Human Out of the Loop: A Review of Bayesian Optimization. *Proc. IEEE* 104, 1 (2016), 148–175. https://doi.org/10.1109/JPROC.2015.2494218

[84] Reza Shokri and Vitaly Shmatikov. 2015. Privacy-Preserving Deep Learning. In *Proceedings of the 22nd ACM SIGSAC Conference on Computer and Communications Security*. Denver, CO, USA, 1310–1321. https://doi.org/10.1145/2810103.2813687

[85] Daniel Stutzbach, Reza Rejaie, Nick Duffield, Subhabrata Sen, and Walter Willinger. 2009. On Unbiased Sampling for Unstructured Peer-to-Peer Networks. *IEEE/ACM Transactions on Networking* 17, 2 (2009), 377–390. https://doi.org/10.1109/TNET.2008.2001730

[86] Zhenheng Tang. 2022. *GossipFL*. Retrieved 2024-08-08 from https://github.com/wizard1203/GossipFL/

[87] Chandra Thapa, Pathum Chamikara Mahawaga Arachchige, Seyit Camtepe, and Lichao Sun. 2022. SplitFed: When Federated Learning Meets Split Learning. *Proceedings of the AAAI Conference on Artificial Intelligence* 36, 8 (2022), 8485–8493. https://doi.org/10.1609/aaai.v36i8.20825

[88] Yongding Tian, Zhuoran Guo, Jiaxuan Zhang, and Zaid Al-Ars. 2023. DFL: High-Performance Blockchain-Based Federated Learning. *Distrib. Ledger Technol.* 2, 3, Article 20 (sep 2023), 25 pages. https://doi.org/10.1145/3600225

[89] Xuezhen Tu, Kun Zhu, Nguyen Cong Luong, Dusit Niyato, et al. 2022. Incentive Mechanisms for Federated Learning: From Economic and Game Theoretic Perspective. *IEEE Transactions on Cognitive Communications and Networking* 8, 3 (2022), 1566–1593. https://doi.org/10.1109/TCCN.2022.3177522

[90] Praneeth Vepakomma, Otkrist Gupta, Tristan Swedish, and Ramesh Raskar. 2018. Split learning for health: Distributed deep learning without sharing raw patient data. https://arxiv.org/abs/1812.00564

[91] Joost Verbraeken, Matthijs Wolting, Jonathan Katzy, Jeroen Kloppenburg, Tim Verbelen, and Jan S. Rellermeyer. 2020. A Survey on Distributed Machine Learning. *Comput. Surveys* 53, 2 (2020), 1–33. https://doi.org/10.1145/3377454

[92] Aidmar Wainakh, Alejandro Sanchez Guinea, Tim Grube, and Max Muhlhauser. 2020. Enhancing Privacy via Hierarchical Federated Learning. In *2020 IEEE European Symposium on Security and Privacy Workshops*. Genoa, Italy, 344–347. https://doi.org/10.1109/EuroSPW51379.2020.00053

[93] Bin Wang, Jun Fang, Hongbin Li, Xiaojun Yuan, and Qing Ling. 2023. Confederated Learning: Federated Learning With Decentralized Edge Servers. *IEEE Transactions on Signal Processing* 71 (2023), 248–263. https://doi.org/10.1109/TSP.2023.3241768

[94] Boyang Wang, Ming Li, Sherman S. M. Chow, and Hui Li. 2014. A tale of two clouds: Computing on data encrypted under multiple keys. In *2014 IEEE Conference on Communications and Network Security*. San Francisco, CA, USA, 337–345. https://doi.org/10.1109/CNS.2014.6997502

[95] Jianyu Wang, Hang Qi, Ankit Singh Rawat, Sashank Reddi, et al. 2022. FedLite: A Scalable Approach for Federated Learning on Resource-constrained Clients. https://arxiv.org/abs/2201.11865

[96] Qian Wang, Minxin Du, Xiuying Chen, Yanjiao Chen, et al. 2018. Privacy-Preserving Collaborative Model Learning: The Case of Word Vector Training. *IEEE Transactions on Knowledge and Data Engineering* 30, 12 (2018), 2381–2393. https://doi.org/10.1109/TKDE.2018.2819673

[97] Stefanie Warnat-Herresthal, Hartmut Schultze, Krishnaprasad L. Shastry, Sathyanarayanan Manamohan, et al. 2021. Swarm Learning for decentralized and confidential clinical machine learning. *Nature* 594, 7862 (2021), 265–270. https://doi.org/10.1038/s41586-021-03583-3

[98] Kang Wei, Jun Li, Ming Ding, Chuan Ma, Howard H. Yang, Farokhi Farhad, Shi Jin, Tony Q. S. Quek, and H. Vincent Poor. 2019. Federated Learning with Differential Privacy: Algorithms and Performance Analysis. https://arxiv.org/abs/1911.00222

[99] Kang Wei, Jun Li, Chuan Ma, Ming Ding, et al. 2023. Personalized Federated Learning With Differential Privacy and Convergence Guarantee. *IEEE Transactions on Information Forensics and Security* 18 (2023), 4488–4503. https://doi.org/10.1109/TIFS.2023.3293417

[100] Claes Wohlin. 2014. Guidelines for snowballing in systematic literature studies and a replication in software engineering. In *Proceedings of the 18th International Conference on Evaluation and Assessment in Software Engineering*. ACM, London England United Kingdom, 1–10. https://doi.org/10.1145/2601248.2601268

[101] Herbert Woisetschläger, Simon Mertel, Christoph Krönke, Ruben Mayer, and Hans-Arno Jacobsen. 2024. Federated Learning and AI Regulation in the European Union: Who is Responsible? – An Interdisciplinary Analysis. http://arxiv.org/abs/2407.08105

[102] Michael Wooldridge, Nicholas R. Jennings, and David Kinny. 2000. The Gaia Methodology for Agent-Oriented Analysis and Design. *Autonomous Agents and Multi-Agent Systems* 3, 3 (2000), 285–312. https://doi.org/10.1023/A:1010071910869

[103] Huiyu Wu and Diego Klabjan. 2023. Robust softmax aggregation on blockchain based federated learning with convergence guarantee. http://arxiv.org/abs/2311.07027

[104] Tengxi Xia, Yongheng Deng, Sheng Yue, Junyi He, et al. 2022. HSFL: An Efficient Split Federated Learning Framework via Hierarchical Organization. In *2022 18th International Conference on Network and Service Management*. Thessaloniki, Greece, 1–9. https://doi.org/10.23919/CNSM55787.2022.9964646

[105] Xun Xian, Xinran Wang, Jie Ding, and Reza Ghanadan. 2020. Assisted Learning: A Framework for Multi-Organization Learning. In *34th Conference on Neural Information Processing Systems*, Vol. 33. Vancouver, Canada, 14580–14591.

[106] Hong Xing, Osvaldo Simeone, and Suzhi Bi. 2020. Decentralized Federated Learning via SGD over Wireless D2D Networks. In *2020 IEEE 21st International Workshop on Signal Processing Advances in Wireless Communications*. Atlanta, GA, USA, 1–5. https://doi.org/10.1109/SPAWC48557.2020.9154332

[107] Minghui Xu, Zongrui Zou, Ye Cheng, Qin Hu, et al. 2023. SPDL: A Blockchain-Enabled Secure and Privacy-Preserving Decentralized Learning System. *IEEE Trans. Comput.* 72, 2 (2023), 548–558. https://doi.org/10.1109/TC.2022.3169436

[108] Rongxu Xu, Wenquan Jin, Anam Nawaz Khan, Sunhwan Lim, and Do-Hyeun Kim. 2023. Cooperative Swarm Learning for Distributed Cyclic Edge Intelligent Computing. *Internet of Things* 22 (July 2023), 100783. https://doi.org/10.1016/j.iot.2023.100783

[109] Xiyuan Yang, Wenke Huang, and Mang Ye. 2023. Dynamic Personalized Federated Learning with Adaptive Differential Privacy. In *Advances in Neural Information Processing Systems*, Vol. 36. Curran Associates, Inc., 72181–72192.

[110] Hao Ye, Le Liang, and Geoffrey Ye Li. 2022. Decentralized Federated Learning With Unreliable Communications. *IEEE Journal of Selected Topics in Signal Processing* 16, 3 (2022), 487–500. https://doi.org/10.1109/JSTSP.2022.3152445

[111] Fangchao Yu, Bo Zeng, Kai Zhao, Zhi Pang, and Lina Wang. 2024. Chronic Poisoning: Backdoor Attack against Split Learning. In *Proceedings of the AAAI Conference on Artificial Intelligence*, Vol. 38. 16531–16538.

[112] Xiao-Tong Yuan and Ping Li. 2020. On convergence of distributed approximate Newton methods: globalization, sharper bounds and beyond. *J. Mach. Learn. Res.* 21, 1 (2020). Publisher: JMLR.org.

[113] Ye Yuan, Jun Liu, Dou Jin, Zuogong Yue, et al. 2023. DeceFL: A Principled Fully Decentralized Federated Learning Framework. *National Science Open* 2, 1 (2023), 1–17. https://doi.org/10.1360/nso/20220043

[114] Yufeng Zhan, Jie Zhang, Zicong Hong, Leijie Wu, et al. 2021. A Survey of Incentive Mechanism Design for Federated Learning. *IEEE Transactions on Emerging Topics in Computing* 10, 2 (2021), 1–1. https://doi.org/10.1109/TETC.2021.3063517

[115] Chengliang Zhang, Suyi Li, Junzhe Xia, Wei Wang, Feng Yan, and Yang Liu. 2020. BatchCrypt: Efficient Homomorphic Encryption for Cross-Silo Federated Learning. In *2020 USENIX Annual Technical Conference*. Boston, MA, USA, 493–506.

[116] Chengliang Zhang, Junzhe Xia, Baichen Yang, Huancheng Puyang, et al. 2021. Citadel: Protecting Data Privacy and Model Confidentiality for Collaborative Learning. In *Proceedings of the ACM Symposium on Cloud Computing*. Seattle, WA, USA, 546–561. https://doi.org/10.1145/3472883.3486998

[117] Tuo Zhang, Lei Gao, Chaoyang He, Mi Zhang, et al. 2022. Federated Learning for the Internet of Things: Applications, Challenges, and Opportunities. *IEEE Internet of Things Magazine* 5, 1 (2022), 24–29. https://doi.org/10.1109/IOTM.004.2100182

[118] Yunfeng Zhao, Zhicheng Liu, Chao Qiu, Xiaofei Wang, et al. 2021. An Incentive Mechanism for Big Data Trading in End-Edge-Cloud Hierarchical Federated Learning. In *2021 IEEE Global Communications Conference*. Madrid, Spain, 1–6. https://doi.org/10.1109/GLOBECOM46510.2021.9685514

[119] Yang Zhao, Jun Zhao, Linshan Jiang, Rui Tan, Dusit Niyato, Zengxiang Li, Lingjuan Lyu, and Yingbo Liu. 2021. Privacy-Preserving Blockchain-Based Federated Learning for IoT Devices. http://arxiv.org/abs/1906.10893

[120] Tianchen Zhou, Zhanyi Hu, Bingzhe Wu, and Cen Chen. 2023. SLPerf: a Unified Framework for Benchmarking Split Learning. https://doi.org/10.48550/arXiv.2304.01502

[121] Hangyu Zhu, Jinjin Xu, Shiqing Liu, and Yaochu Jin. 2021. Federated learning on non-IID data: A survey. *Neurocomputing* 465 (2021), 371–390. https://doi.org/10.1016/j.neucom.2021.07.098

[122] Alexander Ziller, Andrew Trask, Antonio Lopardo, Benjamin Szymkow, et al. 2021. PySyft: A Library for Easy Federated Learning. In *Federated Learning Systems*. Vol. 965. Springer International Publishing, Cham, 111–139. https://doi.org/10.1007/978-3-030-70604-3_5